\documentclass[journal]{IEEEtran}
\usepackage{cite}
\usepackage{graphicx}
\usepackage{amsmath}
\usepackage{algorithmic}
\usepackage{array}
\usepackage{blindtext}
\usepackage{epstopdf}
\usepackage{bm}
\usepackage{enumerate}
\usepackage{amsbsy}
\usepackage{amssymb}
\usepackage{amsthm}
\usepackage{amscd}
\usepackage{subfigure}
\usepackage{color}
\usepackage{booktabs}
\usepackage{multirow}
\usepackage{hhline}
\usepackage{makecell}
\usepackage{bbold}
\usepackage{url}
\usepackage{tabularx}
\usepackage{cellspace}
\usepackage{boldline}
\usepackage{balance}
\usepackage[]{algorithm2e}
\usepackage{dsfont}

\def \bb           {\boldsymbol{b}}
\def \bc           {\boldsymbol{c}}
\def \bd           {\boldsymbol{d}}

\def \br           {\boldsymbol{r}}
\def \bs           {\boldsymbol{s}}
\def \bt           {\boldsymbol{t}}
\def \bu           {\boldsymbol{u}}
\def \bv           {\boldsymbol{v}}

\def \bx           {\boldsymbol{x}}
\def \by           {\boldsymbol{y}}
\def \bz           {\boldsymbol{z}}
\def \bPhi 		 {\boldsymbol{\Phi}}
\def \btau		{\boldsymbol{\tau}}
\def \relu		{\text{ReLU}}
\def \bUpsilon {\boldsymbol{\Upsilon}}
\def \sign 	{\text{sign}}

\def \bR           {\boldsymbol{R}}

\def \bdelta       {\boldsymbol{\delta}}

\def \btau         {\boldsymbol{\tau}}

\def \bUpsilon     {\boldsymbol{\Upsilon}}

\def \bPhi         {\boldsymbol{\Phi}}

\def \bzero        {\boldsymbol{0}}
\def \bone         {\boldsymbol{1}}

\def \calD         {\mathcal{D}}

\def \calG         {\mathcal{G}}

\def \calN         {\mathcal{N}}

\def \calR         {\mathcal{R}}

\def \arg          {\mathrm{arg}}

\def \st           {\mathrm{s.t.}}

\begin{document}

\title{Model-Aware Deep Architectures for One-Bit Compressive Variational Autoencoding}

\author{Shahin~Khobahi$^{\star}$,~\IEEEmembership{Student Member,~IEEE,}
        and~Mojtaba~Soltanalian,~\IEEEmembership{Member,~IEEE} 
        \thanks{This work was supported in part by U.S. National Science Foundation
        	Grant CCF-1704401. Parts of this work
        	have been submitted to the IEEE International Conference on Acoustics, Speech and
        	Signal Processing (ICASSP), Barcelona, Spain, May 2020 \cite{confversion}.
        	\par S. Khobahi and M. Soltanalian are with the Department of Electrical
        and Computer Engineering, University of Illinois at Chicago, Chicago, IL
        60607 (\emph{$^\star$Corresponding author: skhoba2@uic.edu}).}
}

\maketitle

\begin{abstract}
Parameterized mathematical models play a central role in understanding and design of complex information systems. However, they often cannot take into
account the intricate interactions innate to such systems. On the contrary, purely data-driven approaches do not need explicit mathematical models for data generation and have a wider applicability at the cost of interpretability. In this paper, we consider the design of a one-bit compressive variational autoencoder, and propose a novel hybrid model-based and data-driven methodology that allows us not only to design the sensing matrix and the quantization thresholds for one-bit data acquisition, but also allows for learning the latent-parameters of iterative optimization algorithms specifically designed for the problem of one-bit sparse signal recovery. In addition, the proposed method has the ability to adaptively learn the proper quantization thresholds, paving the way for amplitude recovery in one-bit compressive sensing. Our results demonstrate a significant improvement compared to state-of-the-art model-based algorithms.
\end{abstract}


\section{Introduction} \label{sec:intro}
In the past two decades, compressive sensing (CS) has shown significant potential in enhancing sensing and recovery performance in signal processing, occasionally with simpler hardware, and thus, has attracted noteworthy attention among researchers. CS is a method of signal acquisition which ensures the exact or almost exact reconstruction of certain classes of signals using far less number of samples than what is needed in the Nyquist sampling regime \cite{4472240,eldar2012compressed}---where the signals are typically reconstructed by finding the sparsest solution of an under-determined system of equations using various available means.

In a practical setting, each measurement is to be digitized into finite-precision values for further processing and storage purposes, which inevitably introduces a quantization error.
This error is generally dealt with as measurement noise possessing limited energy; an approach that does not perform well in extreme cases.
One-bit CS is one such extreme case where the quantizer is a simple sign comparator and each measurement is represented using only one bit information $r\in\{\pm 1\}$ \cite{ 4558487, 5955138, 6418031, 6404739, 6178284, zhang2014efficient}. 
One-bit quantizers are not only low-cost and low-power hardware components, but also much faster than traditional scalar quantizers, accompanied by great reduction in the complexity of hardware implementation.
Several algorithms have been introduced in the literature for efficient reconstruction of sparse signals in one-bit CS scenarios (e.g., see \cite{4558487, 5955138, 6418031, 6404739, 6178284,zhang2014efficient, 8747470} and the references therein). A detailed discussion of such algorithms is provided in Sec II.
\nocite{khobahi2018signal,ameri2019one,roth2018comparison,zahabi2019one}

\underline{\emph{Notation:}} We use bold lowercase letters for vectors and bold uppercase letters for matrices. $(\cdot)^T$, and $(\cdot)^H$ denote the vector/matrix transpose, and the Hermitian transpose, respectively. $\mathbf{1}$ and $\mathbf{0}$ are the all-one and all-zero vectors. $\|\bx\|_n$ denotes the $\ell_n$-norm of the vector $\bx$ defined as $\left( \Sigma_k|\bx(k)|^n  \right)^{\frac{1}{n}}$. $[\bx]_i$ denotes the $i$-th element of the vector $\bx$. $\text{Diag}(\bx)$ denotes the diagonal matrix formed by the entries of the vector argument $\bx$. The operator $\succeq$ denotes the element-wise vector inequality operator.
\subsection{Relevant Prior Art}
One-bit compressive sensing is mainly concerned with the following data-acquisition model:
\begin{equation}
	\br = \text{sign} (\bPhi\bx - \bb),
\end{equation}
where $\bx\in\mathds{R}^n$ denotes a $K$-sparse source signal, $\bPhi\in\mathds{R}^{m\times n}$ is the sensing matrix, and $\bb\in\mathds{R}^{n}$ denotes the quantization thresholds vector. In addition to the mentioned advantages of using one-bit ADCs for data-acquisition purposes, the use of one-bit information offers increased robustness to undesirable non-linearities in the data-acquition process. Furthermore, there exists strong empirical evidence that recovering a sparse source signal from only one-bit measurement can outperform its multi-bit CS counterpart \cite{6418031,knudson2016one}.

The current one-bit CS recovery algorithms typically exploit the \textit{consistency principle}, which represents the fact that the element-wise product of the sparse signal and the corresponding measurement is always positive \cite{4558487}, i.e. $\br\odot(\bPhi\bx - \bb)\succeq 0$.
However, most of the existing literature on one-bit CS considers zero-level one-bit quantization thresholds (i.e., $\bb=\mathbf{0}$) leading to a total loss of amplitude information during the data-acquisition process. Hence, by comparing the signal level with zero, one can only recover the direction of the source signal, i.e. $\bx/\|\bx\|_2$, and not the amplitude information $\bx$.
In its most general form, any solution $\bx^*$ to the one-bit CS problem should: (i) satisfy the sparsity condition, i.e. $\|\bx^*\|_0\leq K$ with $K = \|\bx\|_0$, and (ii) achieve consistency, i.e. $\br\odot(\bPhi\bx^* - \bb)\succeq 0$. As mentioned above, most of the existing literature on the problem of one-bit CS recovery problem considers the case of $\bb=\mathbf{0}$. In such a case, the solution to the one-bit CS problem can be expressed as:
\[
\bx^*= \underset{\bx}{\text{argmin}}\;\;\|\bx\|_0\;\;\text{s.t.} \;\; \br = \sign(\bPhi\bx).
\]
The above program is NP-hard and mathematically intractable \cite{6418031}. However, there exist several powerful iterative algorithms to find $\bx^*$ (for the case of $\bb=\mathbf{0}$) that rely on a relaxation of the $\ell_0$-norm to its convex hull (i.e., using $\ell_1$-norm in lieu of $\ell_0$-norm) to obtain an estimate of the support of the true source signal by restricting the feasible solutions to the unit-sphere, i.e. $\|\bx\|_2= 1$.

In \cite{4558487}, the authors assume a zero-level quantization threshold and propose an iterative algorithm called \textit{renormalized fixed point iteration} (RFPI) where a convex barrier function is used to enforce the consistency principle (as a regularization term in the objective function). A detailed analysis of the RFPI algorithm is provided in Sec. II. It is worth mentioning that in a traditional CS setting, one consider the under-sampled measurements (i.e., $m<n$), however, the \emph{over-sampling} regime is beneficial and of paramount interest in a one-bit CS setting in that the use of one-bit ADCs provide a cheap and fast way to acquire measurements and to potentially go beyond the limitations of the traditional CS methods.

\begin{figure*}
	\footnotesize
	\centering
	\vspace{-.2cm}
	\includegraphics[width=.9\linewidth]{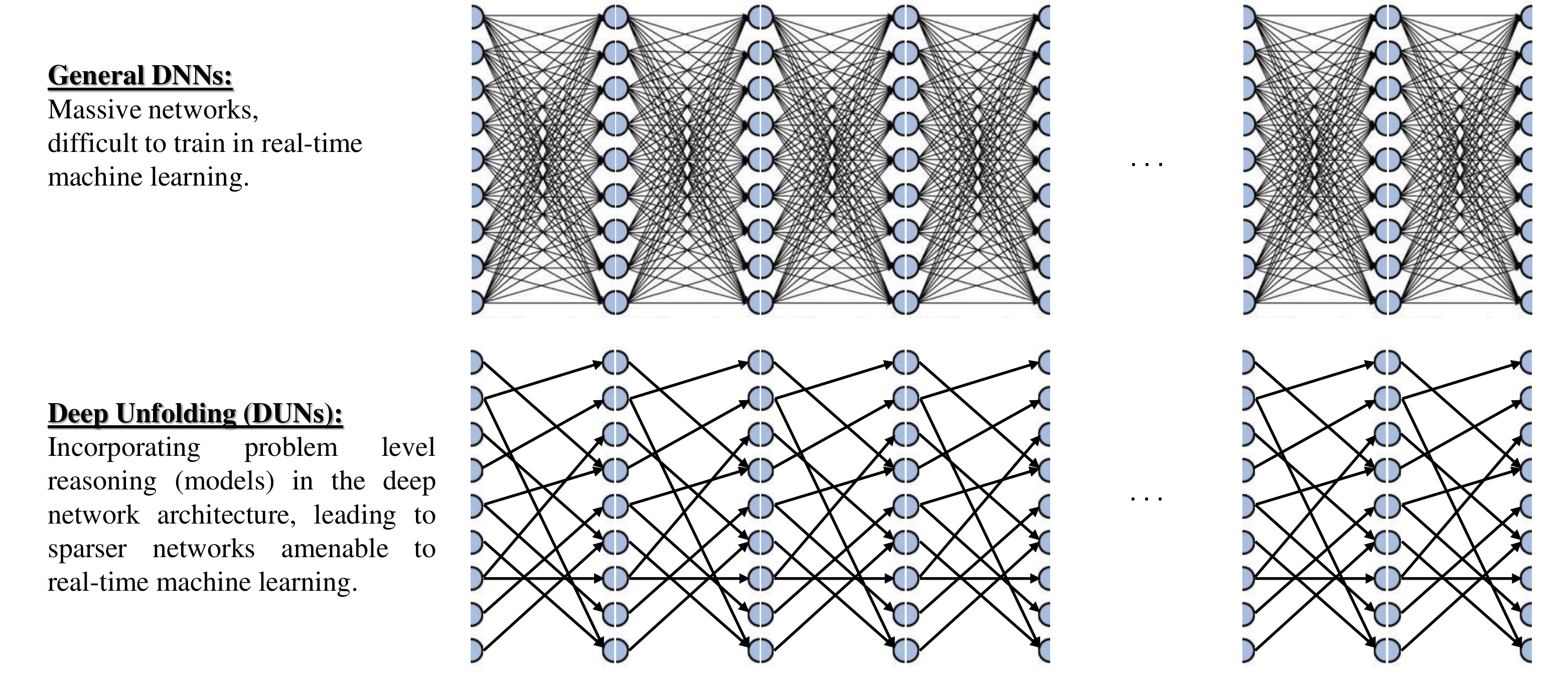}
	\emph{\caption{General DNNs vs DUNs. DUNs appear to be an excellent tool for real-time signal processing and machine learning applications due to the smaller degrees of freedom required for training and execution. \label{fig:deepunfolding}}}
\end{figure*}
Another such reconstruction algorithm can be found in \cite{5955138}, referred to as \textit{restricted step shrinkage} (RSS), for which a nonlinear barrier function is used as the regularizer to enforce the consistency principle. 
Compared to RFPI algorithm, RSS has three important advantages: provable convergence, improved consistency, and feasible performance~\cite{Li2018}.
Ref. \cite{6418031} introduces a penalty-based robust recovery algorithm, called \textit{binary iterative hard thresholding} (BIHT), in order to enforce the consistency principle. Contrary to RFPI algorithm, BIHT exploits the knowledge of the sparsity level of the signal as input, and was shown to be more robust to outliers and have a superior performance than that of the RFPI method in some cases (at the cost of knowing the sparsity level of the source signal a priori).
Both RFPI and BIHT, however, only consider a zero-level quantization threshold, as a result, the amplitude information is lost due to comparing the acquired signal with zero. 
In \cite{6404739} and \cite{6178284}, authors proposed modified versions of RFPI and BIHT, referred to as \textit{noise-adaptive renormalized fixed point iteration} (NARFPI) and \textit{adaptive outlier pursuit with sign flips} (AOP-f), that are more robust against bit flips in the measurement vector (that occur due to the presence of noise). More recently, the authors in \cite{knudson2016one} considered the problem of one-bit CS signal reconstruction in a non-zero quantization thresholds setting that enables the recovery of the norm of the source signal, i.e. recovering $\|\bx\|_2$. However, the proposed method in \cite{knudson2016one} still fails to accurately recover the amplitude information of the source signal, and does not offer a straight-forward apparently to design the quantization thresholds. In addition, there exist several variables in the above mentioned iterative algorithms that must be tuned either heuristically or using expensive computations (e.g., grid-search method) to achieve a high performance. 
In \cite{plan2012robust}, the authors lay the ground work for a theoretical analysis of noisy one-bit CS problem, and propose a novel polynomial-time solver based on a convex programming approach for the problem of one-bit sparse signal recovery in a noisy setting.

Considering the above, it is of paramount importance to develop \emph{computationally efficient} one-bit CS models that can incorporate non-zero quantization thresholds to allow for recovering the amplitude information. Additionally, the vast literature on the one-bit CS recovery problem, does not yet tap into the potential of the available data at hand (to improve the performance recovery). One can significantly benefit from a methodology that can facilitate not only incorporation of the domain knowledge on the problem (i.e., being model-driven), but also the available data at hand to go beyond the performance of the traditional sparsity aware signal processing techniques.

There has recently been a high demand for developing effective real-time signal processing algorithms that use the data to achieve improved performance \cite{7780424, ILIADIS20189, 7447163, shlezinger2019hardware,liao2019deep, shlezinger2019viterbinet}. In particular, the data-driven approaches relying on deep neural architectures such as convolutional neural networks \cite{7780424}, deep fully connected networks \cite{ILIADIS20189}, stacked denoising autoencoders \cite{7447163}, and generative adversarial networks \cite{wu2019deep} have been studied for sparse signal recovery in generic quantized CS settings. we note that, parameterized mathematical models discussed above play a central role in
understanding and design of large-scale information systems
and signal processing methods. However, they often
cannot take into account the intricate interactions innate to
such systems. On the contrary, purely data-driven approaches,
and specifically deep learning techniques, do not need explicit
mathematical models for data generation and have a wider
applicability at the cost of interpretability. The main advantage
of the deep learning-based approach is that it employs several
non-linear transformations to obtain an abstract representation
of the underlying data. Data-driven approaches, on the other
hand, lack the interpretability and trustability that comes with
model-based signal processing. They are particularly prone
to be questioned further, or at least not fully trusted by
the users, especially in critical applications. Furthermore, the
deterministic deep architectures are generic and it is unclear
how to incorporate the existing knowledge on the problem
in the processing stage. \emph{The advantages associated with both
model-based and data-driven methods show the need for
developing frameworks that bridge the gap between the two
approaches.}

The recent advent of the \emph{deep unfolding framework} \cite{chien2017deep,wisdom2017building,khobahi2019deep,hershey2014deep,wisdom2016deep, solomon2019deep} and the corresponding deep unfoling networks (DUNs) has paved the way for a game-changing fusion of models and well-established signal processing approaches with data-driven architectures.
In this way, we not only exploit the vast amounts of available data, but also integrate the prior knowledge of the system model in the processing stage. Deep unfolding relies on the establishment of an optimization or inference iterative algorithm, whose iterations are then \textit{unfolded} into the layers of a deep network, where each layer is designed to resemble one iteration of the optimization/inference algorithm. The resulting hybrid method benefits from low computational cost (in execution stage) of deep neural networks, and at the same time, from the versatility and reliability of model-based methods; thus, appears to be an excellent tool in real-time signal processing applications due to the smaller degrees of freedom required for training and execution (afforded by integration of the problem-level reasoning, or the \emph{model}, see Fig.~\ref{fig:deepunfolding}). 
A detailed analysis of the deep unfolding methodology for the problem of one-bit CS is provided in Sec. \ref{sec:method}.
\subsection{Contributions of the Paper}
In this paper, we propose a novel hybrid model-based and data-driven methodology (based on DUNs) that addresses the drawbacks of both purely model-based (such as the discussed RFPI and BIHT algorithm) and purely data-driven approaches. The resulting methodology is far less data-hungry and assumes a slight over-parametrization of the system model as opposed to traditional deep learning techniques (with very large number of variables to be learned). In particular, the proposed method seeks to bridge the gap between the data-driven and model-based approaches in the one-bit CS paradigm, and to result in a specialized architecture for the purpose of sparse signal recovery from one-bit measurements. The contribution of this paper can be summarized as follows:\\
$\bullet$ We propose a novel hybrid model-based and data-driven one-bit compressive variational autoencoding (VAE) methodology that can deal with the optimization of the sensing matrix $\bPhi$, the one-bit quantization thresholds $\bb$, and the latent-variables of the decoder module according to the underlying distribution of the source signal. Hence, such a methodology allows for quick adaptation to new data distributions and environments.\\
$\bullet$ To the best of our knowledge, this is the first attempt in the one-bit CS paradigm that allows for joint optimization of the quantization thresholds and sensing matrix, also facilitating the \emph{recovery of the amplitude information of the source signal}. We show that by using the proposed VAEs, one can significantly improve upon existing iterative algorithms and gain much higher accuracy both in terms of recovering the magnitude and the support of the underlying source signal.\\ 
$\bullet$ The proposed methodology exhibits performance that goes beyond the traditional one-bit CS state-of-the-art and allows for designing sensing matrices that are distribution-specific. In conjunction to learning data-specific $\bPhi$, the quantization thresholds can also be learned in a joint manner such that the learned parameters improve the signal reconstruction accuracy and speed.\\
$\bullet$ We propose two generalized optimization algorithms that can be used as standalone algorithms for recovering the amplitude information of the source signal by utilizing non-zero quantization thresholds.

\underline{\emph{Organization of the Paper:}} The remainder of this paper is organized as follows. In Sec. II, we discuss the general problem formulation and system model of the one-bit compressive sensing problem and propose two general algorithms that pave the way for incorporating non-zero quantization thresholds. The proposed one-bit compressive variational autoencoding methodology is presented in Sec. III. The loss function characterization and training method for the proposed VAEs are discussed at the end of Sec. III. In Sec. IV, we investigate the performance of the proposed methods through various numerical simulations and for various scenarios. Finally, Sec. V concludes the paper.\nocite{gianelli2016one, ren2019sinusoidal, pirzadeh2019spectral, pun2009opportunistic, rao2019channel, zahabi2019one, rao2019massive}

\section{System Model and Problem Formulation}\label{sec:prob}
In this paper, we are interested in a one-bit CS measurement model (i.e., the encoder module) with dynamics that can be described as follows:
\begin{eqnarray}
\label{eq:7}
\text{Encoder Module: }\quad\quad\br = \sign(\bPhi\bx - \bb),
\end{eqnarray}
where $\bPhi^{m\times n}$ denotes the sensing matrix, $\bb\in\mathds{R}^{m}$ is the quantization thresholds, and $\bx\in\mathds{R}^n$ is assumed to be a $K$-sparse signal. Having the one-bit measurements of the form \eqref{eq:7}, one can pose the problem of sparse signal recovery from one-bit measurements $\br$ by solving the following non-convex program:
\begin{eqnarray}
\mathcal{P}_0:\;\;\underset{\bx}{\min}\;\; \|\bx\|_0,\label{eq:8}\;\;\st \;\;\br = \sign(\bPhi\bx - \bb), \label{eq:9}
\end{eqnarray}
where the constraint in \eqref{eq:9} is imposed to ensure a consistent reconstruction with the available one-bit information. Further note that the one-bit measurement consistency principle in \eqref{eq:9} can be equivalently expressed as 
\begin{eqnarray}
	\label{eq:4}
	\bR \left(\bPhi \bx -\bb \right) \succeq \bzero, 
\end{eqnarray}
where $\bR = \text{Diag}(\br)$. 

Let us first consider the scenario in which the quantization thresholds $\bb$ are all set to zero. In this case, the non-convex optimization problem $\mathcal{P}_0$ can be further relaxed and expressed as a well-known non-convex $\ell_1$-minimization program on the unit sphere \cite{4558487}:
\begin{eqnarray}
\mathcal{P}_1:\;\;\underset{\bx}{\min} &\|\bx\|_1,\;\;
\st \;\;\bR \bPhi \bx \succeq \bzero,\label{eq:11}\;\;
\|\bx\|_2 = 1,\label{eq:12}
\end{eqnarray}
where the $\ell_1$-norm acts as a sparsity inducing function. The intuition behind finding the sparsest signal on the $\ell_2$ unit-sphere (i.e., fixing the energy of the recovered signal) is two-fold. First, it reduces the feasible set of the optimization problem as the amplitude information is lost, and second, it avoids the the trivial solution of $\hat{\bx} = \bzero$. By comparing the acquired data $\by$ with non-zero quantization thresholds, the constraint defined in \eqref{eq:4} not only reduces the feasible set of the problem by defining a set of hyper-planes where the signal can reside on, but also, implicitly exclude the trivial solution. There exists an extensive body of research on approximately solving the non-convex optimization problem $\mathcal{P}_1$ (e.g., see \cite{4558487, 5955138, 6404739, Plan2013, 6638799, zhang2014efficient}, and the references therein). The most notable methods utilize a regularization term $\calR(\bs)$ to enforce the consistency principle via a penalty term added to the $\ell_1$-objective function, viz.
\begin{eqnarray}
\label{eq:6}
\hat{\bx} = \underset{\bx}{\arg\min}\;  \|\bx\|_1  + \alpha\calR(\bR\bPhi\bx),\;
\st	\;\|\bx\|_2 = 1,
\end{eqnarray}
where $\alpha>0$ is the penalty factor.

Among the numerous iterative algorithms available for tackling the optimization problem in \eqref{eq:6}, we plan to utilize and improve upon the state-of-the-art renormalized fixed-point iterations (RFPI) \cite{4558487}, and the Binary Iterative Hard Thresholding (BIHT) \cite{6418031} algorithms as the starting point for our proposed model-driven one-bit compressive variational autoencoding methodology. Namely, in the subsequent sections, we use the mentioned algorithms as a base-line to design the decoder function of our one-bit CS VAE. In particular, we unfold the iterations of the two specialized algorithms onto the layers of a deep neural network in a fashion that each layer of the proposed deep architecture mimics the behavior of one iteration of the base-line algorithm. Next, we perform an end-to-end learning approach by utilizing the back-propagation method to tune the parameters of both the decoder and the encoder functions of the proposed one-bit compressive VAE.

\subsection{Renormalized Fixed-Point Iteration (RFPI)}
The RFPI algorithm considers a one-bit CS data acquisition model where the quantization thresholds are all set to zero. With $\bc = \bR\bPhi\bx$ and $\bb=\mathbf{0}$, the RFPI algorithm utilizes the following regularization term to enforce the consistency constraint in \eqref{eq:11}:
\begin{eqnarray}
\mathcal{R}(\bc) = \frac{1}{2}\,\|\rho(\bc)\|_2^2,
\end{eqnarray}
where $\rho(\bc) \triangleq \max\{-\bc,\bzero\}$, and the function $\max$ is applied element-wise on the vector arguments. Note that the function $\rho(\cdot)$ can be expressed in terms of the well-known  Rectifier Linear Unit ($\relu$) function extensively used by the deep learning research community, i.e. $\rho(\bc) = \relu(-\bc)$. Briefly speaking, the RFPI algorithm is a first-order optimization method (gradient-based) that operates as follows: given an initial point $\bx_0$ on the unit-sphere (i.e., $\|\bx_0\|_2=1$), the gradient step-size $\delta$ and a shrinkage thresholds $\alpha$ (or equivalently the penalty term), at each iteration $i$, the estimated signal $\bx_{i}$ is obtained using the following update steps:
\begin{subequations}
	\label{eq:16s}
	\begin{align}
	\label{eq:16}
	&\bd_i = {\nabla}_{\bx} \, \calR(\bz)\bigr\rvert_{\bx = \bx_{i-1}} = -\left(\bR\bPhi\right)^T\rho\left(\bR\bPhi\bx_{i-1}\right),\\
	\label{eq:17}
	&\bt_i = \left(1 + \delta\bd_i^T\bx_{i-1}\right)\bx_{i-1} - \delta\bd_i,\\
	\label{eq:18}
	&\bv_i = \sign\left(\bt_i\right)\odot\relu\left(|\bt_i| - (\delta/\alpha)\bone\right),\\
	\label{eq:19}
	&\bx_i = \frac{\bv_i}{\|\bv_i\|_2}.
	\end{align}
\end{subequations}
After the descent in \eqref{eq:16}-\eqref{eq:17}, the update step in \eqref{eq:18} corresponds to a shrinkage step. More precisely, any element of the vector $\bt_i$ that is below the threshold $\delta/\alpha$ will be pulled down to zero (leading to enhanced sparsity). Finally, the algorithm projects the obtained vector $\bv_i$ on the unit sphere to produce the latest estimation of the signal. Note that the latter step is necessary due to the fact that a zero-threshold vector (i.e., $\bb=\mathbf{0}$) is employed at the time of the data acquisition, and hence, the amplitude information is lost.

While effective in signal reconstruction, 
there exist several drawbacks in using the RFPI method. For instance, it is required to use the algorithm on several problem instances, while increasing the value of the penalty factor $\alpha$ at each outer iteration of the algorithm, and to use the previously obtained solution as the initial point for tackling the recovery problem for any new problem instance. Moreover, it is not straight-forward how to choose the fixed step-size and the shrinkage threshold, that may depend on the latent-parameters of the system. In fact, it is evident that by carefully tuning the step-sizes and the shrinkage threshold $\tau = \delta/\alpha$, one can significantly boost the performance of the algorithm, and further alleviate the mentioned drawbacks of this method. In what follows, 
we extend the above iterations in a fashion that it allows for incorporating the non-zero quantization thresholds, and hence, enabling us to effectively recover the amplitude information of the source signal.

\textbf{A.1. Extending the RFPI framework to non-zero quantization thresholds:}\\
Recall that our focus is on the following encoding (measurement) model with an arbitrary threshold vector $\bb$:
\begin{eqnarray}
\label{eqq:11}
\br = \sign(\bPhi\bx - \bb).
\end{eqnarray}
Therefore, the problem of one-bit CS signal recovery with a non-zero quantization threshold vector can be cast as:
\begin{eqnarray}
\label{eqq:12}
\underset{\bx}{\min} \;\;\|\bx\|_1,\;\;
\st\;\; \bR (\bPhi \bx - \bb)\succeq \bzero.
\end{eqnarray}
Inspired by the regularization-based relaxation employed in \eqref{eq:6}, we relax the above program and cast it as follows:
\begin{eqnarray}
\mathcal{P}_2:\;\; \underset{\bx}{\min}\; \|\bx\|_1 + \frac{\alpha}{2}\|\rho\left(\bR\left(\bPhi\bx - \bb\right)\right)\|_2^2.
\end{eqnarray}
The above optimization program can be solved in an iterative manner using slightly modified RFP iterations previously described in \eqref{eq:16s}. The slight change presents itself in calculating the gradient of the regularization term, to account for the new measurement model with non-zero thresholds, as well as the exclusion of the projection step onto unit-sphere \eqref{eq:19}. Accordingly, we propose the following new update steps at iteration $i$:

\begin{subequations}
	\underline{\text{The Proposed Generalized RFP Iterations:}}
\label{eqq:15s}
\begin{align}
\label{eqq:15}
&\tilde{\bd}_i = -\left(\bR\bPhi\right)^T\rho\left(\bR\left(\bPhi\bx_{i-1} - \bb\right)\right),\\
\label{eqq:16}
&\tilde{\bt}_i = \bx_{i-1} - \delta\tilde{\bd}_i,\\
\label{eqq:17}
&\bx_i = \sign\left(\tilde{\bt}_i\right)\odot\relu\left(|\tilde{\bt}_i| - (\delta/\alpha)\bone\right).
\end{align}
\end{subequations}
Note that in \eqref{eq:18} there exist an additional projection of the gradient onto the unit sphere through the term $(\bd_i^T\bx_{i-1})\bx_{i-1}$. However, by incorporating the non-zero thresholds vector, such a step is no longer required for the proposed generalized RFP iterations.
In the rest of this paper, we refer to the iterations presented in \eqref{eqq:15s} as \emph{Generalized RFPI} (G-RFPI).
\subsection{Binary Iterative Hard Thresholding Algorithm (BIHT)}
The BIHT algorithm is a simple, yet powerful, first-order iterative reconstruction algorithm for the problem of one-bit CS where the sparsity level $K$ is assumed to be known a priori. BIHT iterations can be seen as a simple modification of the iterative hard thresholding (IHT) algorithm proposed in \cite{blumensath2009iterative}. Similar to the RFPI algorithm, the BIHT method considers a zero-level quantization threshold. However, in contrast to the RFPI algorithm, it exploits the knowledge of the sparsity level $K$ of the signal of interest. In other words, the BIHT algorithm is designed to tackle the following counterpart of $\mathcal{P}_0$:
\begin{eqnarray}
\label{eqs:16}
	\mathcal{P}_3:\;\;
	\underset{\bx}{\min} \;\;\alpha\|\rho\left(\bR\bPhi\bx\right)\|_1,\;\;
	\label{eqs:17}
	\st\;\;	\|\bx\|_0 = K,\;\;
	\label{eqs:18}
	\|\bx\|_2 = 1,
\end{eqnarray}
where $\rho(\bc) = \max\{-\bc, \mathbf{0}\}$ and $\bR = \text{Diag}(\br)$ as before. Note that the one-sided $\ell_1$ objective function above (also related to the hinge-loss) enforces the consistency principle previously introduced in \eqref{eq:11}, and that by solving the above optimization problem, we are working to achieve \emph{maximal consistency} with the one-bit measurements $\br$. It is worth mentioning that one can also consider different objective functions, and not necessarily an  $\ell_1$ objective, as long as it promotes the data consistency principle (e.g., $\ell_2$ norm). For a detailed analysis of different candidates for the objective function and their properties, see \cite{blumensath2009iterative}.

The BIHT iterations are described as follows. Let $\bc = \bR\bPhi\bx$, and define $\mathcal{F}(\bc) = \|\rho(\bc)\|_1$. Given an initial point $\bx_0$, the sparsity level $K$, and one-bit measurements $\br$ (or equivalently $\bR$), at the $i$-th iteration, the BIHT algorithm updates the current estimate of the signal $\bx_{i-1}$ through the following steps:
\begin{subequations}
	\label{eqqs:19}
\begin{eqnarray}
	&\bu_{i} &= \bx_{i-1} - \frac{\alpha}{2}\partial\mathcal{F}(\bx_{i-1})\nonumber\\
	&& = \bx_{i-1} + \frac{\alpha}{2}\bPhi^T\left(\br - \sign(\bPhi\bx_{i-1})\right),\label{eqs:19}\\
	&\bx_{i} &= \mathcal{H}_{K}\left(\bu_{i}\right)\label{eqs:20},
\end{eqnarray}
\end{subequations}
where $\partial\mathcal{F}$ denotes the sub-gradient of the one-side $\ell_1$ objective function in $\mathcal{P}_3$, $\alpha>0$ governs the fixed gradient step-size, and the projecton operator $\mathcal{H}_K(\bx)$ is defined such that it retains the largest $K$ elements (in magnitude) of the vector argument, and set the rest of the elements to zero.

The step \eqref{eqs:19} can be interpreted as taking a descent step using the computed sub-gradient of the objective function \eqref{eqs:16}, while the projection step in \eqref{eqs:20} can be viewed as a projection of $\bu_i$ onto the support set of $K$-sparse signals. Once the above iterations terminate either by fully satisfying the consistency principle (i.e., obtaining $\bx^*$ such that $\mathcal{F}(\bx^*) = 0$), or by achieving a maximum number of iterations, the ultimate step to be taken is projecting the final estimate $\bx^*$ onto the unit-sphere, viz. $\bx^*\leftarrow \bx^*/\|\bx^*\|_2$. Note that this is in contrast to the RFPI algorithm as the BIHT iterations does not require a normalization step as in \eqref{eq:19} at each iteration.

\vspace{2pt}
\textbf{B.1. Extending the BIHT framework to non-zero quantization thresholds:} 

The extension of the BIHT iterations to incorporate the non-zero thresholds vector $\bb$ is straight-forward. In the case of non-zero quantization thresholds, we cast the signal recovery problem as 
\begin{eqnarray}
\underset{\bx}{\min} \;\;\alpha\|\rho\left(\bR(\bPhi\bx - \bb)\right)\|_1, \;\;\st\;\;	\|\bx\|_0 = K,
\end{eqnarray}
where $\bR = \text{Diag}(\br)$, and $\br = \sign(\bPhi \bx - \bb)$.

Similar to the steps we took in \eqref{eqq:11}-\eqref{eqq:15s}, and by employing some rudimentary algebraic operations, the proposed generalized update steps of the BIHT algorithm may be expressed as:

\begin{subequations}
	\vspace{4pt}
	\underline{\text{The Proposed Generalized BIHT Iterations:}}
	\label{eqq:21s}
	\begin{eqnarray}
	&\bu_{i} & = \bx_{i-1} + \frac{\alpha}{2}\bPhi^T\left(\br - \sign(\bPhi\bx_{i-1} - \bb)\right),\label{eqq:21a}\\
	&\bx_{i} &= \mathcal{H}_{K}\left(\bu_{i}\right)\label{eqq:21b},
	\end{eqnarray}
\end{subequations}
with the exception that in the proposed generalized BIHT iterations, there is no need for the normalization of the obtained estimate of the signal $\bx^*$ after the update steps terminate. This is due to the fact that a non-zero quantization threshold vector is employed at time of the encoding, and hence, the amplitude information is not fully lost. In the rest of this paper, we refer to the above iterations as \emph{Generalized BIHT} (G-BIHT) algorithm.

Although simple and powerful, the BIHT algorithm requires a careful choice of the gradient step-size $\alpha$ for convergence, and there is no straight-forward method to properly choose the gradient step-size. On the other hand, it only utilizes a fixed step-size along all iterations. Hence, this motivates the development of a methodology through which one can design a decoder function that exploits adaptive gradient step-sizes, i.e. by considering a different step-size at each iteration, that can result in a significant improvement of the performance of the BIHT algorithm. 

In the next section, we discuss a slight over-parametrization of the iterations of RFPI, G-RFPI, BIHT, and G-BIHT algorithms that paves the way for the design of our proposed one-bit compressive VAE and for jointly designing the parameters of the encoder function defined in \eqref{eq:7} parametrized on the sensing matrix $\bPhi$, the quanitzation thresholds $\bb$, and the design of a set of decoder functions based on the discussed iterative optimization algorithms.

\section{The Proposed One-Bit Compressive Variational Autoencoding Approach}\label{sec:method}
We pursue the design of a novel \emph{model-driven one-bit compressive sensing-based variational autoencoder} deep architecture that facilitates the joint design of the parameters of both the encoder and the decoder module when one-bit quantizers with non-zero thresholds are employed in the data acquisition process (i.e., the encoding module) for a $K$-sparse input signal $\bx\in\mathds{R}^n$. 

In general terms, a variational AE is a generative model comprised of an encoder and a decoder module that are sequentially connected together. The purpose of an AE is to learn an abstract representation of the input data, while providing a powerful data reconstruction system through the decoder module. The input to such a system is a set of signals following a certain distribution, i.e. $\bx\sim\mathcal{D}(\bx)$, and the output is the recovered signal from the decoder module $\hat{\bx}$. Hence, the goal is to jointly learn an abstract representation of the underlying distribution of the signals through the encoder module, and simultaneously, learning a decoder module allowing for reconstruction of the compressed signals from the obtained abstract representations. Therefore, an AE can be defined by two main functions: \emph{i}) an encoder function $f^{\text{Encoder}}_{\bUpsilon_1}: \mathds{R}^n \mapsto \mathds{R}^m$, parameterized on a set of variables $\bUpsilon_1$ that maps the input signal into a new vector space, and \emph{ii}) a decoder function $f^{\text{Decoder}}_{\bUpsilon_2}: \mathds{R}^m \mapsto \mathds{R}^n$ parameterized on $\bUpsilon_2$, which maps the output of the encoder module back into the original signal space. Hence, the governing dynamics of a general VAE can be expressed as
\begin{eqnarray}
\hat{\bx} =  f^{\text{Decoder}}_{\boldsymbol{\Upsilon}_2} \circ f^{\text{Encoder}}_{\boldsymbol{\Upsilon}_1}(\bx),
\end{eqnarray}
where $\hat{\bx}$ denotes the reconstructed signal.

In light of the above, we seek to interpret a one-bit CS system as an VAE module facilitating not only the design of the sensing matrix $\bPhi$ and the quantization thresholds $\bb$ that best captures the information of a $K$-sparse signal when one-bit quantizers are employed, but also to learn the parameters of an iterative optimization algorithm specifically designed for the task of signal recovery. To this end, we modify and unfold the iterations of the proposed G-RFPI algorithm defined in \eqref{eqq:15s}, and the GBIHT method defined in \eqref{eqq:21s} onto the layers of a deep neural network and later use the deep learning tools to tune the parameters of the proposed one-bit compressive VAE.
\subsection{Structure of the Encoding Module}
\begin{figure}
\centering
	\includegraphics[width=0.4\textwidth]{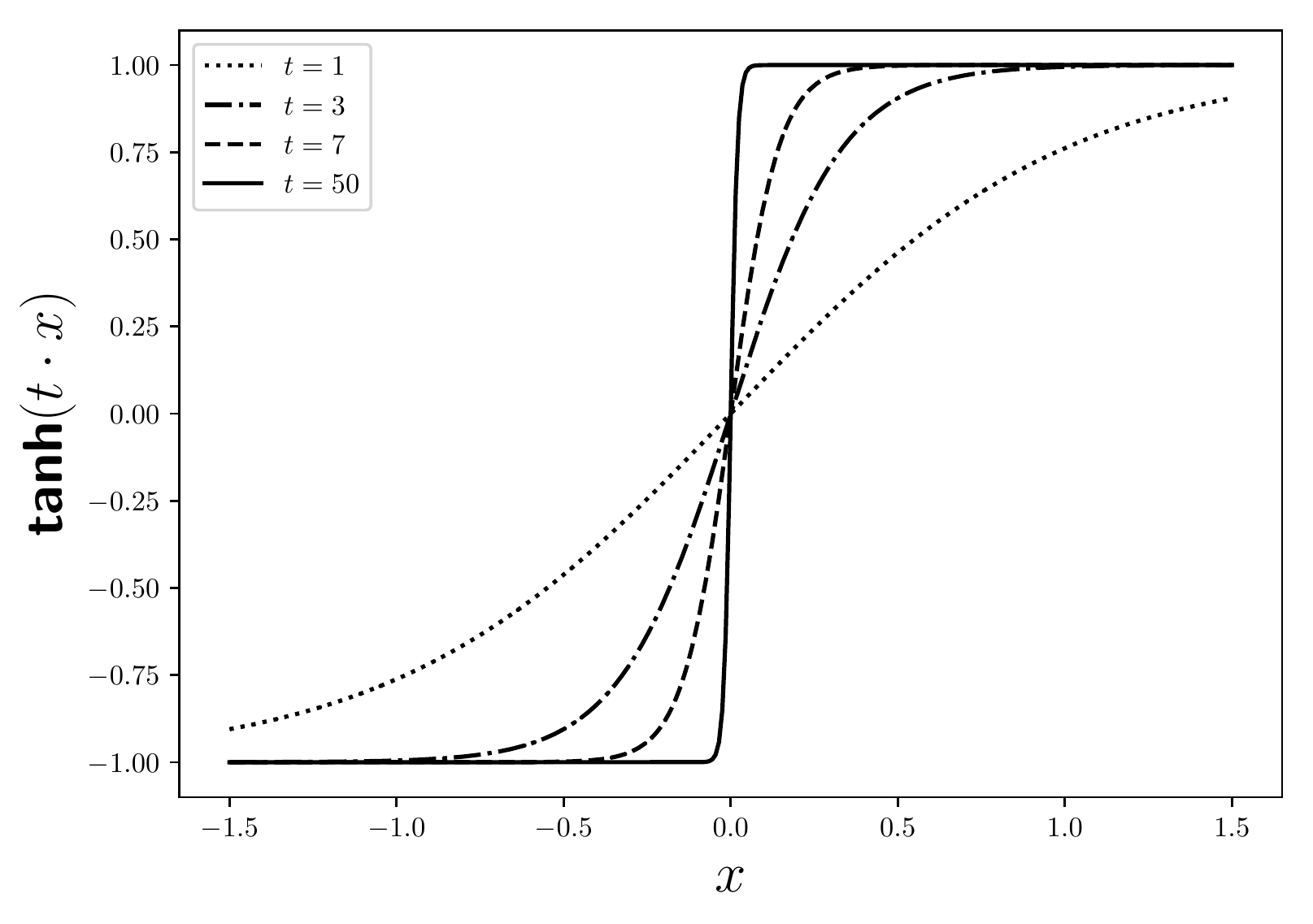}
\caption{\small The function $\tanh(t\cdot x)$ as an approximation of the $\sign$ function for $t\in\{1,3,7,50\}$.}
\label{fig:tanh}
\end{figure}

In its most general form, we define the encoder module of the proposed VAE based on our data-acquisition model defined in \eqref{eq:7}, as follows:
\begin{eqnarray}
\label{eq:20}
f^{\text{Encoder}}_{\bUpsilon_1}(\bx) = \tilde{\sign}(\bPhi\bx -\bb),
\end{eqnarray}
where $\bUpsilon_1=\{\bPhi, \bb\}$ denotes the set of learnable parameters of the encoder function, and  $\tilde{\sign}(\bx) = \tanh(t\cdot\bx)$, for a large $t>0$ ($t$ was set to $50$ in numerical investigations). Note that we replaced the original $\sign$ function with a smooth differentiable approximation of it based on the hyperbolic tangent function. The reason for such a replacement is that the $\sign$ function is not continuous and its gradient is zero everywhere except at the origin, and hence, the use of it would cripple stochastic gradient-based optimization methods (later used in back-propagation method for deep learning). 
Fig. \ref{fig:tanh} plots the function $\tilde{\sign}(x)$ for $t\in\{1,3,7,50\}$, also demonstrating that larger values of $t$ allow for better approximations of the original $sign$ function. 
\subsection{Structure of the Decoding Module}
In this part, we describe the different scenarios under which we pursue the design of our decoder function by using the RFPI, BIHT, and the suggested G-RFPI and G-BIHT iterations. In particular, we fix the total complexity of our decoding module by fixing the total number of iterations allowed for the mentioned optimization iterations. Next, we slightly over-parameterize each iteration/step of the mentioned algorithms to increase the per-iteration degrees-of-freedom of each method and to further account for the learnable latent variables in the system. Finally, we unfold the iterations of each algorithm onto the layers of a deep architecture such that each layer of the deep network resembles one iteration of the base-line algorithm. We then seek to learn the parameters of both the decoder and encoder function using the training tools already developed for deep learning. We consider the following cases to design our decoder function:\\ 
$\bullet$ \textbf{Learned~RFPI~(L-RFPI)}: We consider the RFPI iterations defined in \eqref{eq:16s} as our base-line but slightly over-parametrize its iterations by introducing a gradient step-size $\delta_i$ and a \emph{shrinkage} thresholds vector $\btau_i$ \emph{for each iteration} $i$. This is in contrast to the original RFP iterations where a fixed gradient step-size $\delta$, and shrinkage threshold $\tau = (\delta/\alpha)\mathbf{1}$ were employed for all iterations. Hence, the proposed unfolded over-parametrized iterations are much more expressive. The decoder function will be parameterized on $\bUpsilon_2=\{\delta_i,\btau_i\}_{i=0}^{L-1}$, and the encoder function will be parametrized on the set $\bUpsilon_1=\{\bPhi\}$ (note that $\bb=\mathbf{0}$).\\ 
$\bullet$ \textbf{Learned BIHT (L-BIHT)}: We consider the unfolding of the iterations of the BIHT defined in \eqref{eqqs:19} similar to the previous case and by introducing per-iteration gradient step-sizes $\delta_i$ in lieu of a fixed gradient-step size along all iterations. In this case, the decoder function will be parametrized on the set $\bUpsilon_2=\{\delta_i  \}$, while the set of parameters of the encoding module is $\bUpsilon_1=\{\bPhi\}$; both are to be learned.\\
$\bullet$ \textbf{Learned~G-RFPI~(LG-RFPI)}: We consider the unfolding of the proposed Generalized RFPI iterations in \eqref{eqq:15s} in a non-zero quantization thresholds setting. We over-parameterize the iterations of the proposed G-RFPI by parametrizing the decoder function on the set $\bUpsilon_2=\{\delta_i, \btau_i\}_{i=0}^{L-1}$, and this time, by parameterizing the encoder function on both the sensing matrix and the quantization thresholds vector, i.e. $\bUpsilon_1 = \{\bPhi, \bb\}$.\\
$\bullet$ \textbf{Learned~G-BIHT~(LG-BIHT)}: We consider the unfolding of the G-BIHT iterations defined in \eqref{eqq:21s} in a similar manner, i.e. by parameterizing the decoder function on $\bUpsilon_2 = \{\delta_i\}_{i=0}^{L-1}$. However, similar to the previous case, we further parametrize the encoder function on the quantization thresholds vector in conjunction with the sensing matrix, i.e. $\bUpsilon_1 = \{\bPhi,\bb\}$.
\subsection{The Proposed One-Bit Compressive Variational Authoencoding Methodology}
In the following, we describe the design of four novel deep architectures based on the above mentioned structures and discuss the governing dynamics of the proposed one-bit compressive sensing-based VAE.

\textbf{C.1. \underline{L-RFPI-Based Compressive Autoencoding}}:\\
In this case, we consider the following parameterized encoder function:
\begin{eqnarray}
\label{eq:23}
f^{\text{Encoder}}_{\bUpsilon_1}(\bx) = \tilde{\sign}(\bPhi\bx),\;\; \text{where } \bUpsilon_1 = \{  \bPhi  \}. 
\end{eqnarray}
As for the decoder function, and based on the RFPI iterations in \eqref{eq:16s}, define $g_{\phi_i}: \mathds{R}^m \mapsto \mathds{R}^n$ as follows:
\begin{subequations}
	\label{eq:10}
\begin{align}
\label{eq:10a}
	&g_{\phi_i}(\bz;\bPhi,\bR) = \frac{\bv}{\|\bv\|_2},\quad \text{with}\\
\label{eq:10b}
	&\bv = \tilde{\sign}\left(\bt\right)\odot\relu\left(|\bt| - \btau_i\right),\\
	\label{eq:10c}
	&\bt = \left(1 + \delta_i\bd^T\bz\right)\bz - \delta_i\bd,\\
	\label{eq:10d}
	&\bd = -\left(\bR\bPhi\right)^T\rho\left(\bR\bPhi\bz\right),
\end{align}
\end{subequations}
where $\phi_i = \{\btau_i, \delta_i\}$ represents the parameters of the function $g_{\phi_i}$, and $\btau_i\in\mathds{R}^{n}$ denotes the sparsity inducing \emph{shrinkage} thresholds vector, and $\delta_i$ represents the gradient step-size at iteration $i$. Next, we define the proposed L-RFPI composite decoder function as follows:
\begin{eqnarray}
\label{eqq:25}
	f^{\text{Decoder}}_{\bUpsilon_2}(\bz_0) = g_{\phi_{L-1}}\circ g_{\phi_{L-2}}\circ \dots \circ g_{\phi_{1}} \circ g_{\phi_{0}}(\bz_0;\bPhi,\bR),
\end{eqnarray}
where $\bUpsilon_2=\{\phi_i\}_{i=0}^{L-1}$ represents the learnable (tunable) parameters of the decoder function, and $\bz_0$ is some initial point of choice. Note that we have over-parameterized the iterations of the RFPI algorithm by introducing the new variable $\btau_i$ at each iteration for the sparsity inducing step in \eqref{eq:10b}. Moreover, in contrast with the original RFPI iterations, we have introduced a new step-size $\delta_i$ at each step of the iteration as well (see Eq. \eqref{eq:10c}). Therefore, the above decoder function can be interpreted as performing $L$ iterations of the original RFPI algorithm with an additional $L(n+1) - 2$ degrees of freedom (as compared to the base algorithm) expressed in terms of the set of the shrinkage thresholds $\btau_i$ and the gradient step-sizes $\delta_i$, i.e. $\{\btau_i, \delta_i\}_{i=0}^{L-1}$. As a reslt, the proposed decoder function is much more expressive than that of the iterations of RFPI algorithm. 

\underline{\emph{Remark}}: Note that the above encoder and decoder function, once cascaded together, can be viewed as a deep neural network with $(L+1)$ layers, where the dynamics of the first layer is described by the encoder function defined in \eqref{eq:23}, and the governing dynamics of the succeeding layers is described by compuations of the form \eqref{eq:10a}-\eqref{eq:10d}. Equivalently, such a deep architecture can be viewed as a computational graph with shared variables among the computation nodes, and thus, its parameters can be efficiently optimized by utilizing known deep learning tools such as back-propagation. Hence, the goal is to \emph{jointly} learn the parameters of such a cascaded network (i.e., $\bUpsilon_1\cup\bUpsilon_2$) in an end-to-end manner by using the available data at hand coming from the underlying distribution of the source signal $\bx$.\hfill$\blacksquare$

\textbf{C.2. \underline{L-BIHT-Based Compressive Autoencoding:}}\\
Similar to the previous case, we consider the same encoding function parametrized only on the learnable sensing matrix $\bPhi$ in a zero quantization thresholds setting, i.e. $\bUpsilon_1 = \{  \bPhi \}$. The governing equations for the decoder function in the case of the proposed Learned BIHT are as follows. We re-define $g_{\phi_i}: \mathds{R}^m \mapsto \mathds{R}^n$ as:
\begin{subequations}
	\label{eq:26}
	\begin{align}
	\label{eq:26a}
	&g_{\phi_i}(\bz;\bPhi,\br,K) = \mathcal{H}_{K}\left(\bv\right),\quad\text{for } i<L, \text{where}\\
	\label{eq:26b}
	&\bv = \bz + \delta_i\bPhi^T\left(\br - \tilde{\sign}(\bPhi\bz)\right),
	\end{align}
\end{subequations}
with $\phi_i = \{\delta_i\}_{i=0}^{L-1}$, and where we have an added final layer $i=L$, to renormalize the reconstructed signal as, 
\begin{eqnarray}
\label{eqq:27}
g_{\phi_{L}} (\bz;\bPhi, \br) = \frac{\bz}{\|\bz\|_2}.
\end{eqnarray}
Therefore, similar to the previous case, the proposed L-BIHT-based decoder function is defined as:
\begin{eqnarray}
\label{eqq:28}
f^{\text{Decoder}}_{\bUpsilon_2}(\bz_0) = g_{\phi_{L}}\circ g_{\phi_{L-1}}\circ \dots \circ g_{\phi_{1}} \circ g_{\phi_{0}}(\bz_0;\bPhi,\bR, K).
\end{eqnarray}
We again observe the slight over-parametrization of the L-BIHT algorithm during the unfolding process. Namely, at each iteration we are introducing the per-iteration step-sizes $\delta_i$ to be learned which further enhances the performance of our iterations (see \eqref{eq:26}). In this case, the decoder function is parameterized only on the gradient step-sizes, i.e. $\bUpsilon_2 = \{   \delta_i\}_{i=0}^{L-1}$. The L-BIHT iterations have an additional $(L-1)$ degrees of freedom compared to that of the original BIHT iterations.

\vspace{4pt}
\textbf{C.3. \underline{LG-RFPI-Based Compressive Autoencoding:}}\\
We consider the unfolding of iterations of the Learned Generalized RFPI method according to \eqref{eqq:15s}. As previously discussed, in the generalized iterations of both the RFPI and BIHT algorithms, the encoder module can be expressed as:
\begin{eqnarray}
\label{eqq:29}
f^{\text{Encoder}}_{\bUpsilon_1}(\bx) = \tilde{\sign}(\bPhi\bx - \bb),
\end{eqnarray}
where $\bUpsilon_2 = \{ \bPhi, \bb  \}$, and $\bb$ represents the tunable vector of quantization thresholds. We follow a similar approach to the proposed L-RFPI-Based deep architecture and slightly over-parameterize the iterations in \eqref{eqq:15}-\eqref{eqq:17}, leading to the design of the decoder function:
\begin{subequations}
	\label{eq:30}
	\begin{align}
	\label{eq:31a}
	&g_{\phi_i}(\bz;\bPhi,\bR) = \tilde{\sign}\left(\bv\right)\odot\relu\left(|\bv| - \btau_i\right),\quad\text{with}\\	
	\label{eq:31b}
	&\bv = \bz - \delta_i\bd,\\
	\label{eq:31c}
	&\bd = -\left(\bR\bPhi\right)^T\rho\left(\bR\left(\bPhi\bz - \bb\right)\right),
	\end{align}
\end{subequations}
where $\phi_i = \{\btau_i, \delta_i\}$ represents the parameters of the function $g_{\phi_i}$, $\btau_i\in\mathds{R}^{n}$ denotes the sparsity inducing thresholds vector, and $\delta_i$ represents the gradient step-size at iteration $i$. Hence, the proposed decoder function $f^{\text{Decoder}}_{\bUpsilon_2}(\bz_0)$ can be represented in the same way as in \eqref{eqq:25}, with $\bUpsilon_2 = \{\phi_i\}_{i=0}^{L-1}$. Note that by incorporating the non-zero quantization thresholds, there is no need for an additional normalization term at each iteration. The above iterations (comprising the decoder function) have the same degree of freedom as L-RFPI iterations---an additional $L(n+1) - 2$ model parameters compared to that of the base-line G-RFPI iterations. Also, note the additional $m$ degrees of freedom that the encoder function offers in terms of tunable quantization thresholds vector $\bb$ (in addition to the sensing matrix).

\vspace{4pt}
\textbf{C.4. \underline{LG-BIHT-Based Compressive Autoencoding:}}\\
We consider an encoder function $f^{\text{Encoder}}_{\bUpsilon_1}$ of the form \eqref{eqq:29}, where $\bUpsilon_1=\{\bPhi, \bb  \}$ denotes the learnable sensing matrix and arbitrary quantization thresholds. Additionally, we present an over-parameterization of the Genralized BIHT iterations (see Eqs. \eqref{eqq:21s}) and consider the resulting unfolded network as the blueprint of our decoder. Namely, we define $g_{\phi_i}: \mathds{R}^{m}\mapsto \mathds{R}^{n}$ as:
\begin{subequations}
	\label{eq:25}
	\begin{align}
	\label{eq:25a}
	&g_{\phi_i}(\bz;\bPhi,\br, K) = \mathcal{H}_{K}\left(\bv\right),\quad \text{with}\\
	\label{eq:25b}
	&\bv = \bz + \delta_i\bPhi^T\left(\br - \tilde{\sign}(\bPhi\bz - \bb)\right),
	\end{align}
\end{subequations}
where $\phi_i = \{\delta_i\}$ denotes the set of parameters of the function $g_{\phi_i}$. Note that due to employing a non-zero thresholds vector, we do not need the additional normalization layer as in \eqref{eqq:27} for this case. Consequently, the decoder function $f^{\text{Decoder}}_{\bUpsilon_2}$ can be expressed in a similar manner as in \eqref{eqq:28}, with $\bUpsilon_2 = \{\delta_i \}_{i=0}^{L-1}$. These iterations, similar to L-BIHT case, have an additional $(L-1)$ degrees of freedom compared to that of the base-line G-BIHT iterations; whereas, the encoder function has an additional $m$ tunable parameters  in terms of the one-bit quantization thresholds compared to that of the L-BIHT-based AE.

In the next section, we discuss the training process of the above proposed one-bit compressive autoencoders. Particularly, we formulate a proper loss function that facilitates the training of such unfolded deep architectures, and for each model, we seek to jointly learn the set of parameters of the entire network (i.e., the encoder and decoder function) in a end-to-end manner using the available deep learning techniques. 
\subsection{Loss Function Characterization and Training Method}
\label{sec:training}
The output of an autoencoder is the reconstructed signal from the compressed measurements, i.e.
\[
\hat{\bx} =  f^{\text{Decoder}}_{\boldsymbol{\Upsilon}_2} \circ f^{\text{Encoder}}_{\boldsymbol{\Upsilon}_1}(\bx),
\]
where $\bx\sim\mathcal{D}(x)$ and $\hat{\bx}$ denotes the input and output of the AE, respectively. The training of an AE should be carried out by defining a proper loss function $\calG\left(\bx,f^{\text{Decoder}}_{\bUpsilon_2} \circ f^{\text{Encoder}}_{\bUpsilon_1}(\bx)\right)$ that provides a measure of the similarity between the input and the output of the AE. The goal is to minimize the distance between the input target signal $\bx$ and the recovered signal $\hat{\bx}$ according to a similarity criterion. A widely-used option for the loss function is the output MSE loss, i.e., $\mathds{E}_{\bx\sim\calD(\bx)}\left\{\|\bx - \hat{\bx}\|_2^2\right\}$, and hence, the training loss of such a system can be formulated as:
\begin{eqnarray}
	\calG(\bx;\hat{\bx}) = \mathds{E}_{\bx\sim\calD(\bx)}\left\{\|\bx - f^{\text{Decoder}}_{\bUpsilon_2} \circ f^{\text{Encoder}}_{\bUpsilon_1}(\bx)\|_2^2\right\}\nonumber
\end{eqnarray}
that is to be minimized over $\bUpsilon_1$ and $\bUpsilon_2$. Nevertheless, in deep architectures with a high number of layers and parameters, such a simple choice of the loss function makes it difficult to back-propagate the gradients; in fact, the vanishing gradient problem arises. Therefore, for the training of the proposed AE, a better choice for the loss function is to consider the cumulative MSE loss of the layers. As a result, one can also feed-forward the decoder function for only $l<L$ layers (a lower complexity decoding), and consider the output of the $l$-th layer as a good approximation of the target signal. For training, one needs to consider the constraint that the gradient step-sizes $\{\delta_i\}_{i=0}^{L-1}$, and the shrinkage thresholds $\{ \btau_i \}_{i=0}^{L-1}$ must be non-negative. By parameterizing the decoder function on the step-sizes and the shrinkage step thresholds, we need to regularize the training loss function ensuring that the network chooses positive step sizes and shrinkage thresholds at each layer. With this in mind, we suggest the following loss function for training the proposed one-bit compressive AE. Let $\tilde{g}_i = g_{\phi_i} \circ g_{\phi_{i-1}} \circ \dots \circ  g_{\phi_{0}} \circ f_{\bUpsilon_1}^{\text{Encoder}}(\bx)$, and define the loss function for training as
\begin{align}
\label{eqq:32}
\calG_{L}(\bx;\hat{\bx}) = &\underbrace{\sum_{i=0}^{L-1}w_i ||\bx - \tilde{g}_i(\bx_{i})||_2^2}_{\text{accumulated MSE loss of all layers}} + \\
&\underbrace{\lambda\sum_{i=0}^{L-1} \relu(-[\bdelta]_i) + \lambda\sum_{i=0}^{nL-1} \relu(-[\btau]_i)}_{\text{regularization term for the step-sizes and shrinkage thresholds}},
\nonumber
\end{align}
where $w_i$ denotes the importance weight of choice for the output of each layer, $\lambda>0$, $[\bdelta]_i = \delta_i$, and $\btau = [\btau_0^T, \dots,\btau_{L-1}^T]$. Note that as the information flows through the network, one expects that as we progress layer by layer, the reconstruction shows improvement. A reasonable weighting scheme for designing the importance weights $w_i$ is to gradually increase the importance weights as we proceed through the layers. In this work, we consider a logarithmic weighting scheme, i.e. $w_i = \text{log}(i+2), \text{for }i\in\{0,\dots,L-1\}$. Moreover, in training the autoencoders based on the BIHT algorithm, we exclude the last term in \eqref{eqq:32} as there is no shrinkage thresholds required for these models.

As for the training procedure, our numerical investigations showed that an incremental learning approach is most effective for training of the proposed networks. The details of the incremental learning method that we employed are as follows. During the $l$-th increment round (for $l \in \{ 0,\dots,L-1\}$), we seek to optimize the cost function $\mathcal{G}_{l}(\bx,\hat{\bx})$ by learning the set of parameters $\bUpsilon_l = \bUpsilon_1 \cup \{\phi_i\}_{i=0}^{l}$. At each round $l$, we perform a batch learning with mini-batches of size $B$. After finishing the $l$-th round of training, the $(l+1)$-th layer will be added to the network, and the objective function will be changed to $\mathcal{G}_{l+1}(\bx,\hat{\bx})$. Next, the entire network will go through another batch-learning phase. Interestingly, in this method of training, the learned parameters from the $l$-th round $\bUpsilon_l$ will be used as the initial values of the same parameters in the next round. 

\section{Numerical Results}
\begin{figure*}
    \centering
    \subfigure[]{\includegraphics[width=0.24\textwidth]{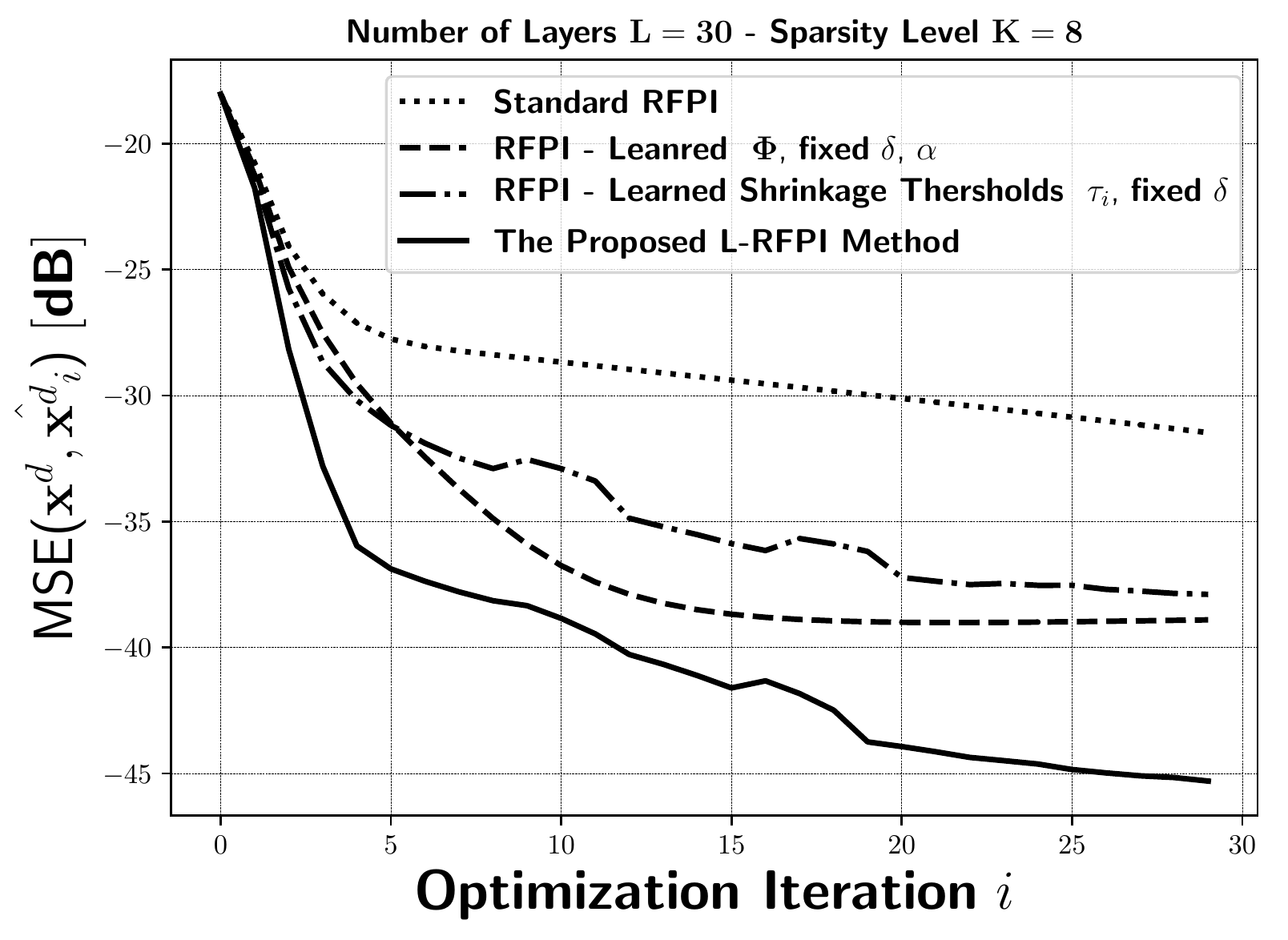}} 
    \subfigure[]{\includegraphics[width=0.24\textwidth]{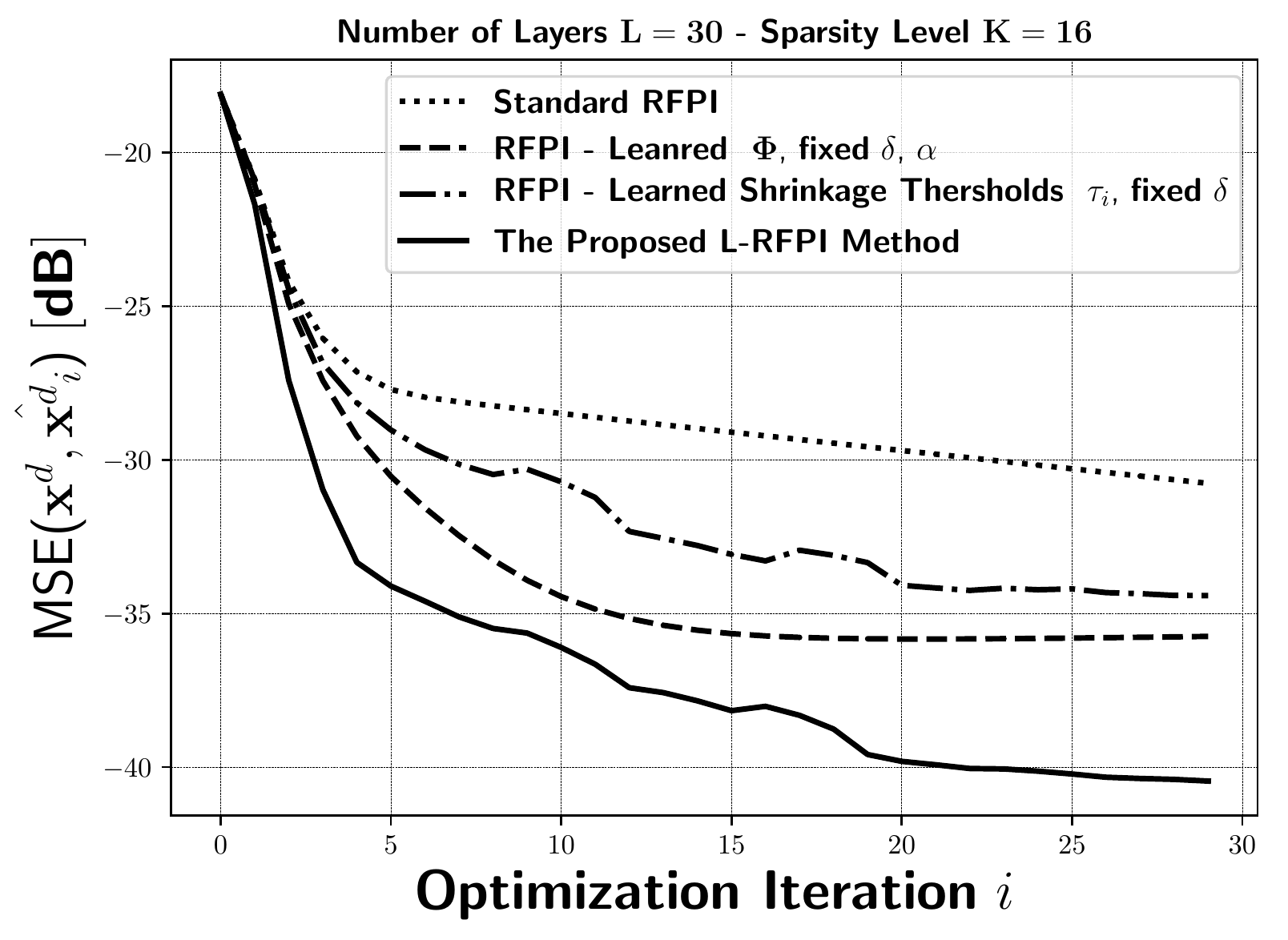}} 
    \subfigure[]{\includegraphics[width=0.24\textwidth]{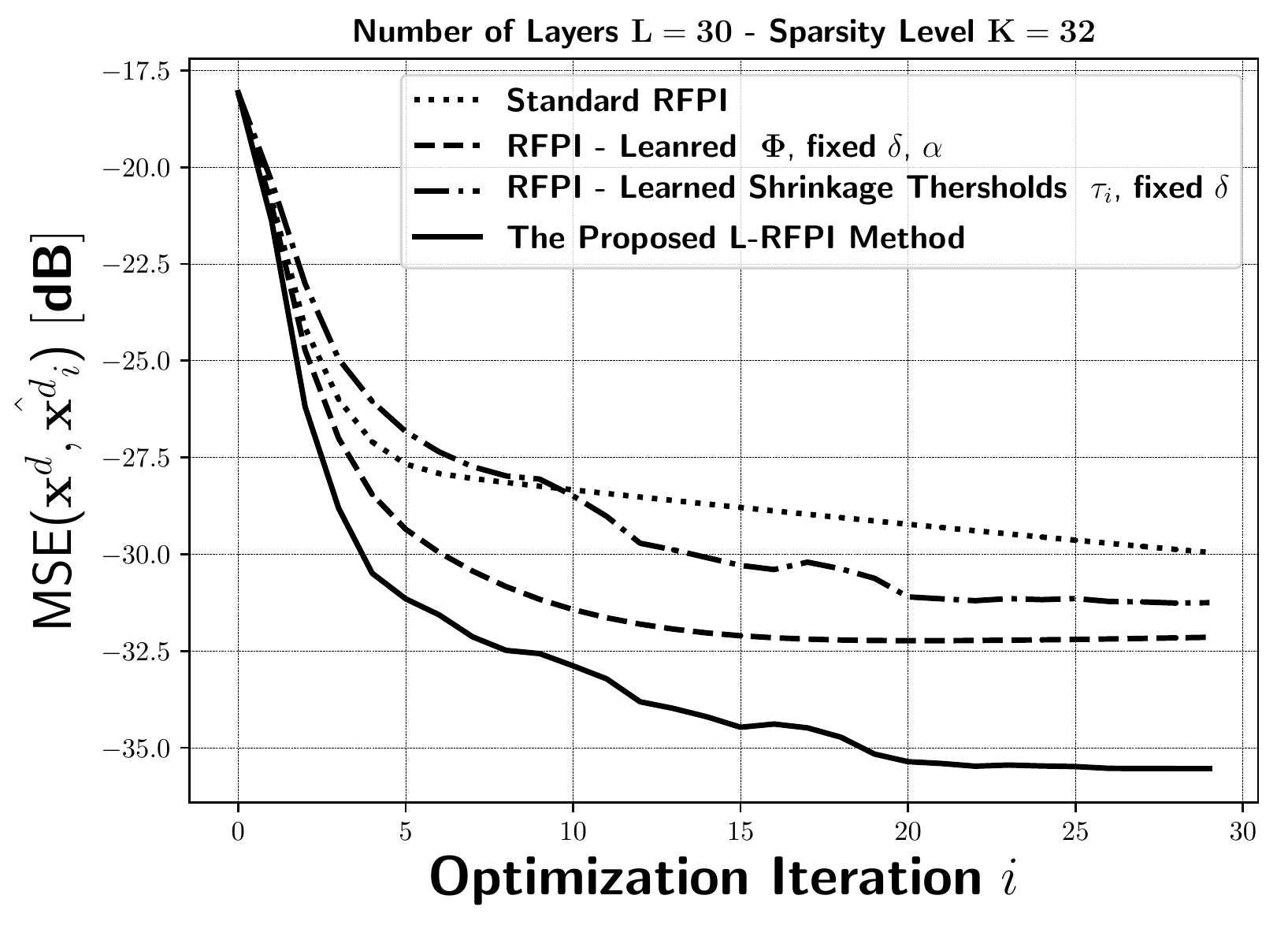}} 
    \subfigure[]{\includegraphics[width=0.24\textwidth]{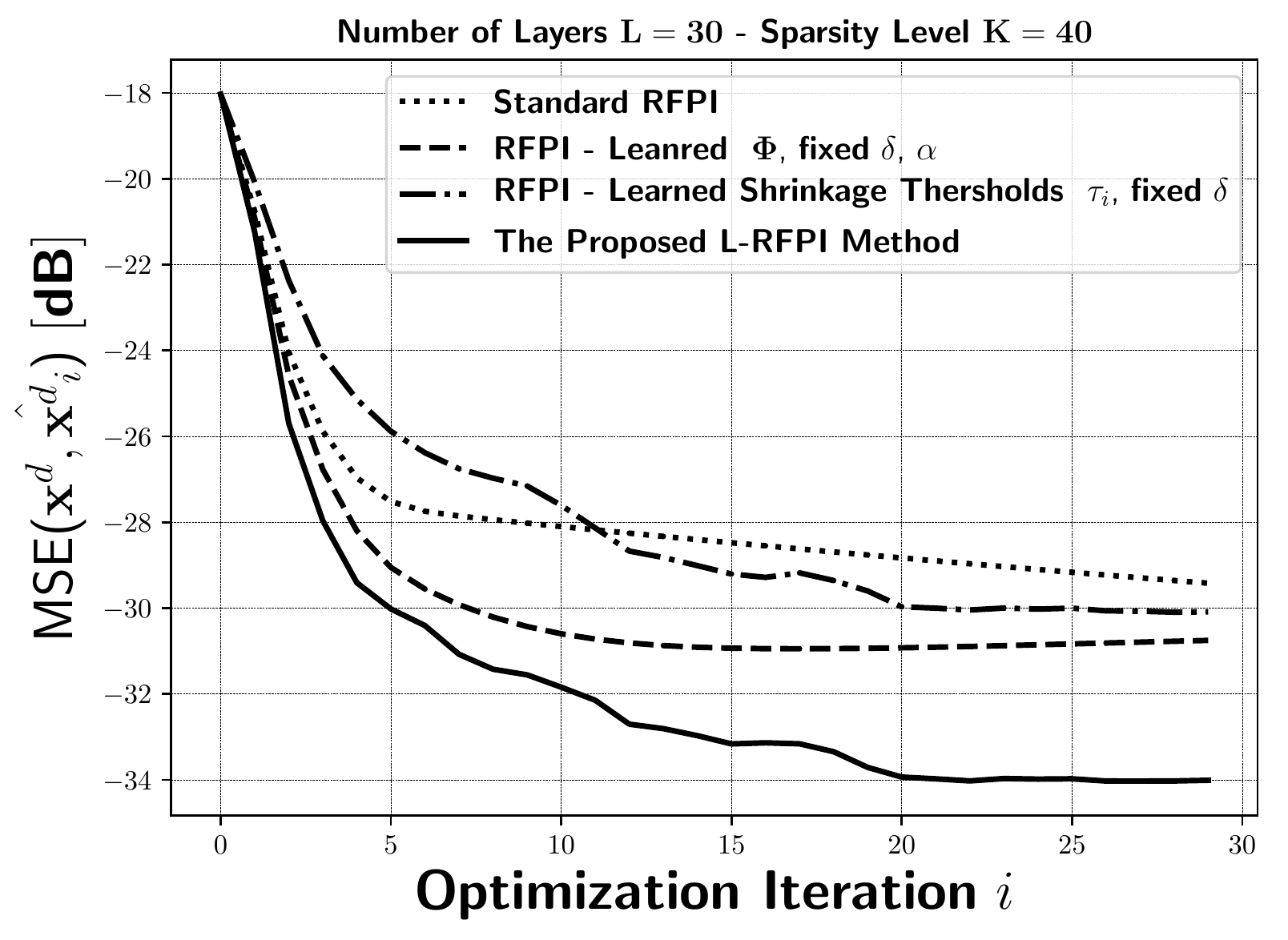}} 
    \caption{The performance of the proposed L-RFPI method compared to the RFPI algorithm for sparsity levels: (a) $K=8$, (b) $K=16$, (c) $K=32$, and (d) $K=40$.}
    \label{fig:2}
\end{figure*}
\begin{figure*}
    \centering
    \subfigure[]{\includegraphics[width=0.24\textwidth]{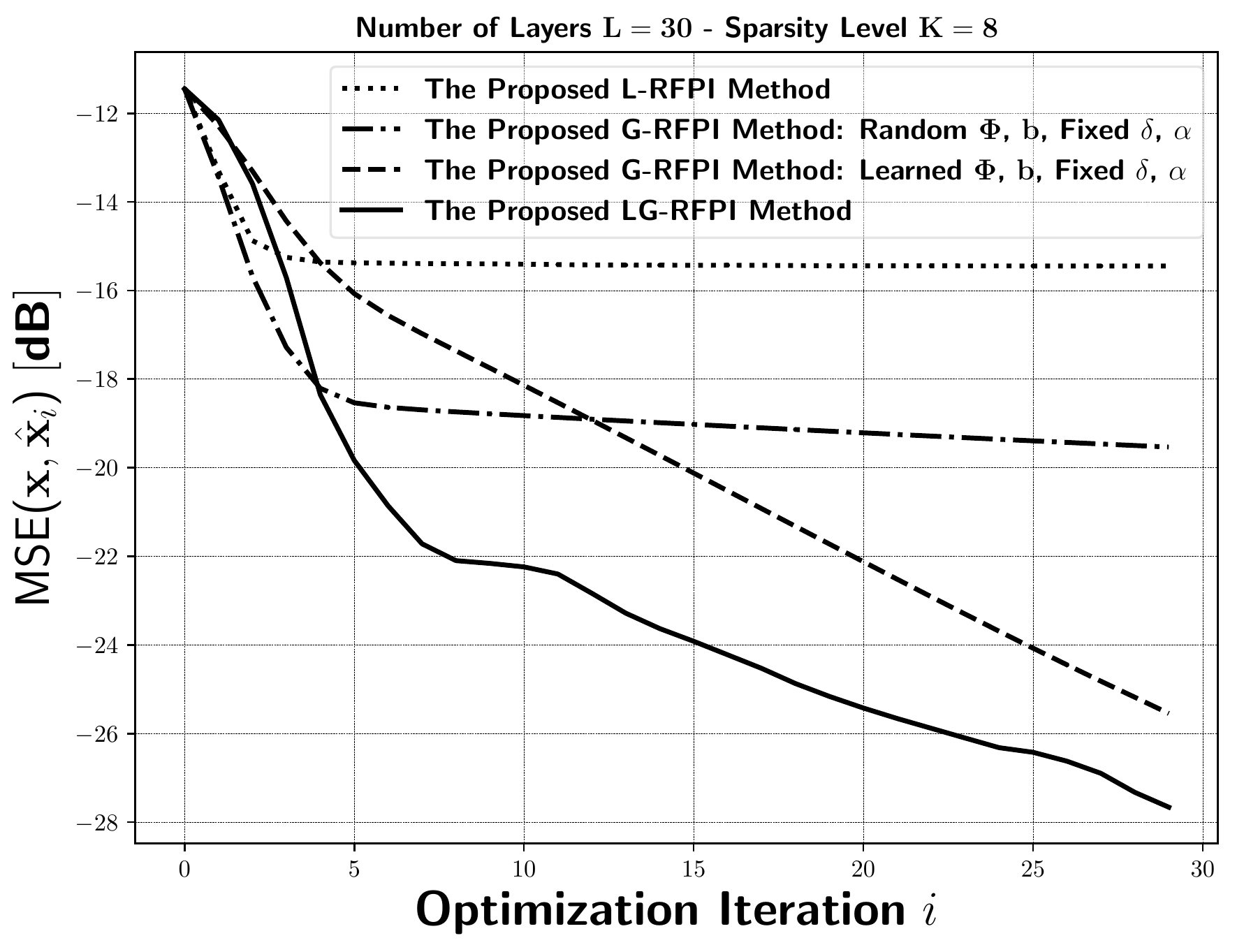}} 
    \subfigure[]{\includegraphics[width=0.24\textwidth]{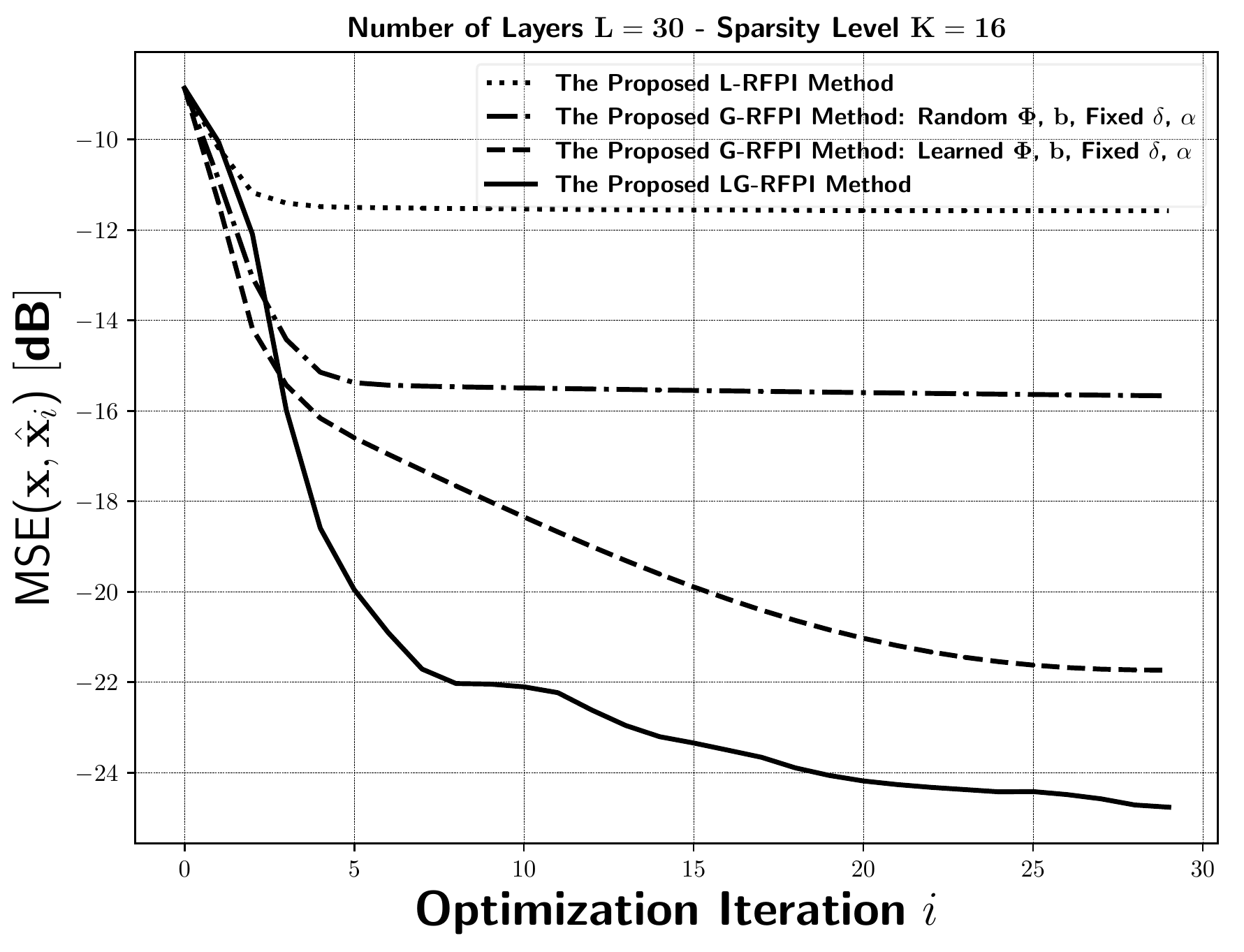}} 
    \subfigure[]{\includegraphics[width=0.24\textwidth]{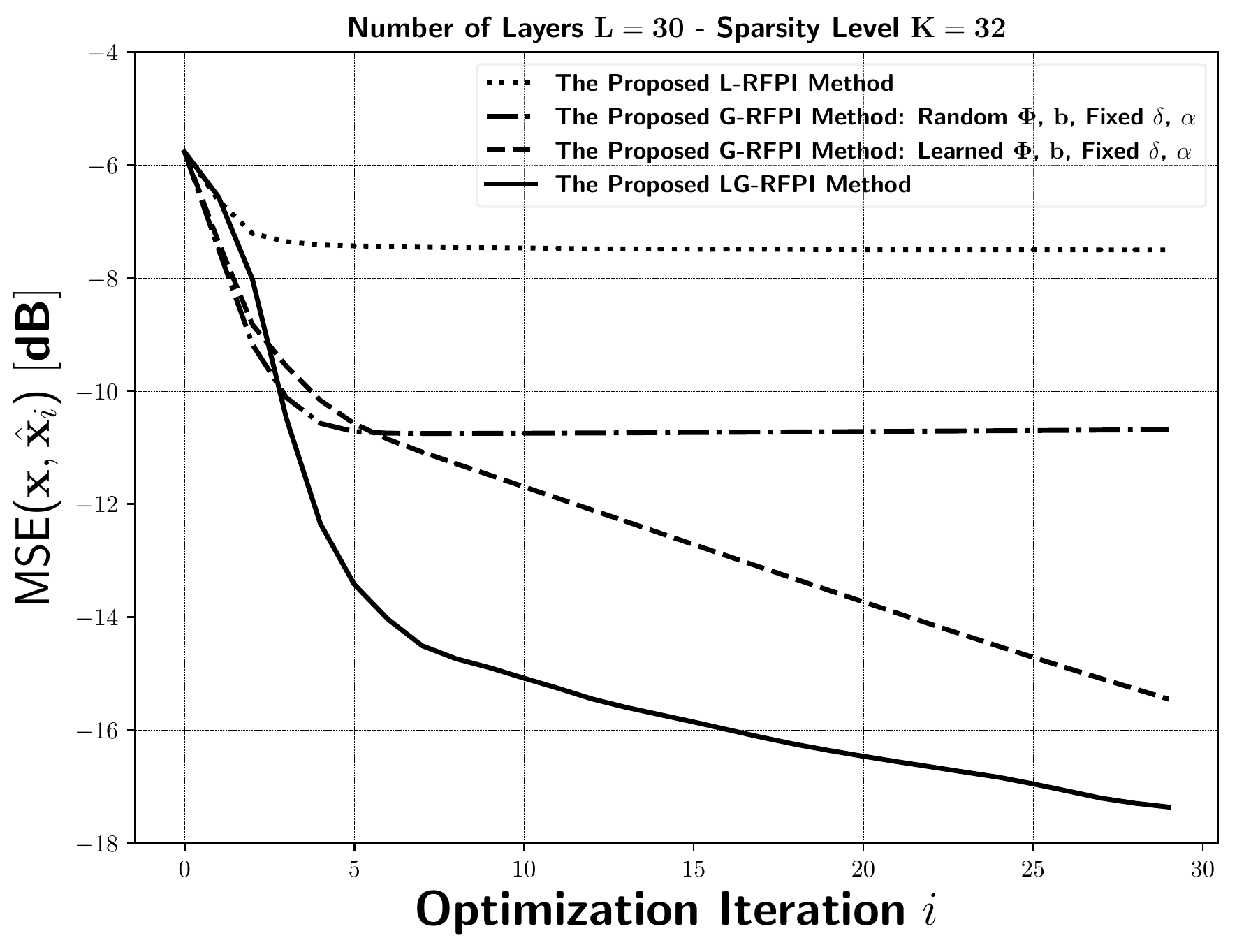}} 
    \subfigure[]{\includegraphics[width=0.24\textwidth]{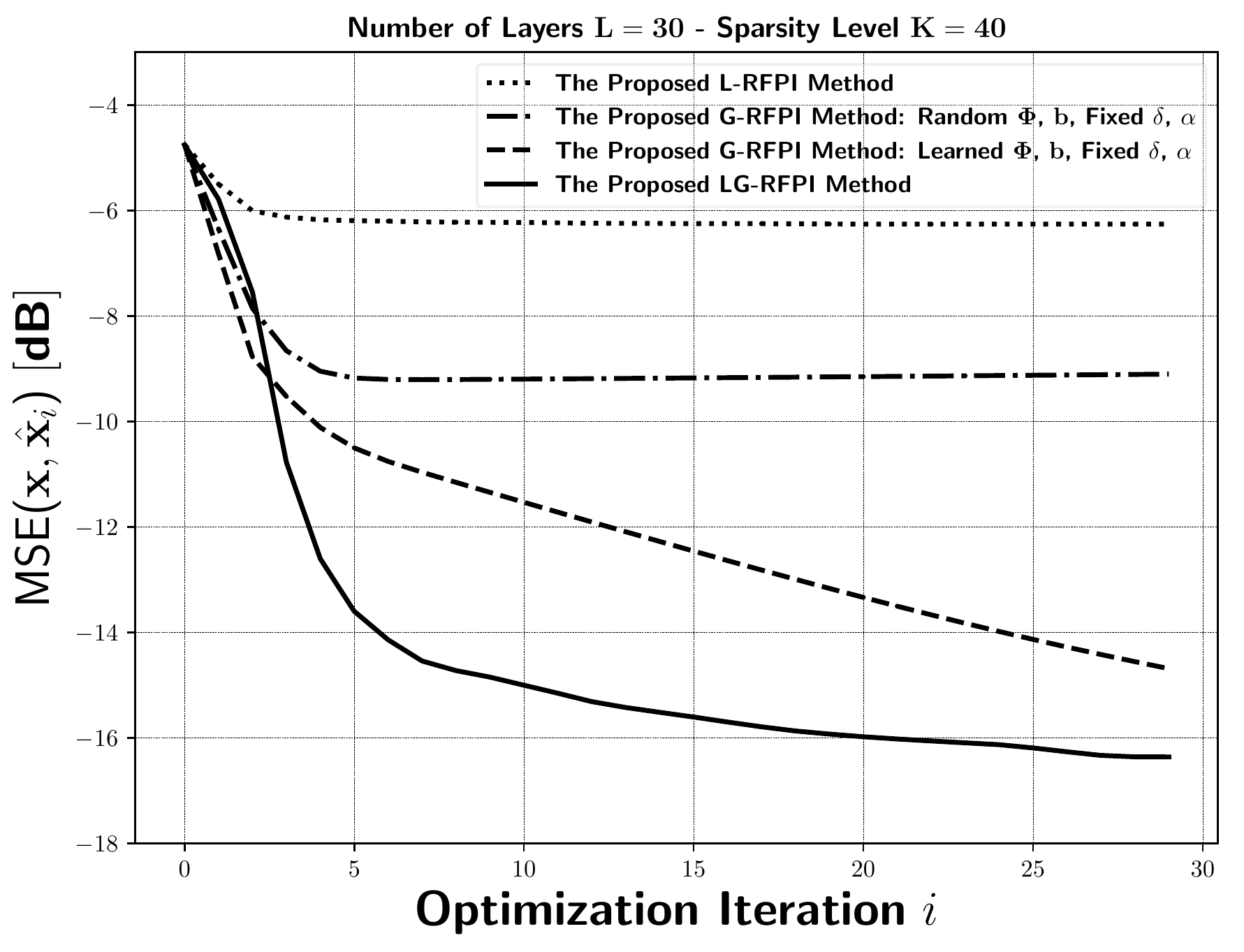}} 
    \caption{The performance of the proposed LG-RFPI VAE and the proposed G-RFPI method in recovering the amplitude information of the $K$- sparse signals for sparsity levels: (a) $K=8$, (b) $K=16$, (c) $K=32$, and (d) $K=40$.}
    \label{fig:3}
\end{figure*}
In this section, we present various simulation results to investigate the performance of the proposed one-bit compressive VAEs and to further show the effectiveness of our training. For training purposes, we randomly generate $K$-sparse signals of length $n=128$, i.e. $\bx\in\mathds{R}^{128}$ where the non-zero elements are sampled from $\mathcal{N}(0,1)$. Furthermore, we fix the total number of layers of the decoder function to $L=30$; equivalent of performing only 30 optimization iterations of the form \eqref{eq:10}, \eqref{eq:26}, \eqref{eq:30}, and \eqref{eq:25}. As for the sensing matrix (to be learned), we assume $\bPhi\in\mathds{R}^{512\times 128}$. The results presented here are averaged over $128$ realizations of the system parameters. Similar to \cite{4558487}, we consider the case that $m>n$, due to the focus of this study on one-bit sampling where usually a large number of one-bit samples are available, as opposed to the usual infinite-precision CS settings.


The proposed one-bit CS VAEs are implemented using the $\mathrm{PyTorch}$ library \cite{paszke2017automatic}. The Adam optimizer \cite{kingma2014adam} with a learning rate of $10^{-3}$ is utilized for optimization of parameters of the proposed deep architectures. Due to the importance of reproducible research, we have made all the codes implemented publicly available along with this paper.\footnote{The code is also available at: \footnotesize{https://github.com/skhobahi/deep1bitVAE}}

As it was previously discussed in Sec. \ref{sec:training}, we employ an incremental batch-learning approach with mini-batches of size $64$ at each round $l<30$, and a total number of $200$ epochs per round. For training of the the proposed VAEs that are based on the RFPI iterations, i.e., the L-RFPI and LG-RFPI deep architectures, we uniformly sample the sparsity level of the source signal from the set $K\in\{16,24,32\}$ for each training point in the mini-batch. We evaluate the performance of the proposed methods on target signals with $K\in\{16,32\}$, as well as $K\in\{8, 40\}$ (which was not presented to the network during the training phase). Moreover, due to the fact that the BIHT method and the corresponding one-bit VAEs (L-BIHT and LG-BIHT) require the knowledge of the sparsity level of the source signal a priori, there is no need to train the network on various sparsity levels; i.e., the corresponding deep architectures can be trained for a particular $K$. Hence, for the L-BIHT and LG-BIHT deep architectures, we train the network for source signals with $K=16$, and evaluate the performace of the resulted networks on $K \in \{ 16, 24 \}$. 

In the sequel, we refer to $\bs^{\text{d}} = \bs/\|\bs\|_2$ as the \emph{normalized} version of the vector $\bs$. In all scenarios, in order to have a fair comparison between the algorithms, the initial starting point $\bz_0$ of the optimization algorithms are the same.

\textbf{\underline{Performance of the proposed L-RFPI VAE}:}\\
In this part, we investigate the performance of the proposed L-RFPI-based VAE in recovering the normalized source signal $\bx$, i.e., recovering $\bx_{i}^\text{d}$.

Fig.~\ref{fig:2} illustrates Mean Squared Error (MSE) for normalized version of the recovered signal $\hat{\bx}_i^\text{d}$ versus total number of optimization iterations $i$, for $i\in\{ 0,\dots,29\}$, and for sparsity levels (a) $K=8$, (b) $K=16$, (c) $K=32$, and (d) $K=40$. We compare the performance of the proposed L-RFPI algorithm with the standard RFPI iterations in \eqref{eq:16}-\eqref{eq:19}, in the following scenarios:\\ $\bullet$ {\emph{Case 1}}: The RFPI algorithm with a randomly generated sensing matrix whose elements are i.i.d. and sampled from $\calN(0,1)$, and fixed values for $\delta$ and $\alpha$.\\ $\bullet$ {\emph{Case 2:}} The RFPI algorithm where the learned $\bPhi$ is utilized, and the values for $\delta$ and $\alpha$ are fixed as in the previous case.\\ $\bullet$ {\emph{Case 3:}} The RFPI algorithm with a randomly generated $\bPhi$ (same as Case 1), however, the learned shrinkage thresholds vector $\{\btau_i\}_{i=0}^{L-1}$ is utilized with a fixed step-size.\\ $\bullet$ {\emph{Case 4:}} The proposed one-bit L-RFPI VAE method corresponding to the iterations of the form \eqref{eq:10a}-\eqref{eq:10d}, with learned $\bPhi$, $\{\delta_i\}_{i=0}^{L-1}$, and $\{\btau_i\}_{i=0}^{L-1}$.

To have a fair comparison, we fine-tuned the parameters of the standard RFPI method (Case 1), i.e., the step-size $\delta$ and the shrinkage threshold $\alpha$, using a grid-search method. It can be seen from Fig.~\ref{fig:2} that in all cases of $K \in \{8, 16, 32, 40\}$, the proposed L-RFPI method demonstrates a significantly better performance than that of the RFPI algorithm (described in Case 1)---an improvement of $\sim 10$ times in MSE outcome. Furthermore, the effectiveness of the learned $\bPhi$ (Case 2), and the learned $\{\btau_i\}$ (Case 3) compared to the base algorithm (Case 1), are clearly evident, as both algorithms with learned parameters significantly outperform the original RFPI. Finally, although we trained the network for $K\in\{16,24,32\}$ sparse signals, it still shows very good generalization properties even for $K\in\{8,40\}$ (see Fig.~\ref{fig:2} (a) and (d)). This is presumably due to the fact that the proposed L-RFPI-based VAE is a \emph{hybrid} model-based data-driven approach that exploits the existing domain knowledge of the problem as well as the available data at hand. Furthermore, note that the proposed method achieves a high accuracy very quickly and does not require solving \eqref{eq:6} for several instances as opposed to the original RFPI algorithm---thus showing great potential for usage in real-time applications.

\noindent
\vspace{2pt}
\textbf{\underline{Performance of the proposed LG-RFPI VAE }:}\\
Next, we investigate the performance of the proposed LG-RFPI VAE (see Eqs. \eqref{eq:31a}-\eqref{eq:31c}) and the G-RFPI algorihtm (see Eqs. \eqref{eqq:15}-\eqref{eqq:17}) that we specifically designed for incorporating arbitrary quantization thresholds at data acquisiton. We investigate the performance of the proposed method in both cases of recovering the amplitude information as well as the normalized signal. 

Fig.~\ref{fig:3} illustrates the MSE between the source signal $\bx$ and the recovered signal $\hat{\bx}_i$ versus total number of optimization iterations $i$, for $i\in\{0,\dots,29\}$, and for sparsity levels (a) $K=8$, (b) $K=16$, (c) $K=32$, and (d) $K=40$. Similar to the previous case, we consider the following scenarios:\\
$\bullet$ {\emph{Case 1}}: The proposed G-RFPI algorithm with a randomly generated sensing matrix and quantization threhsolds vector, whose elements are i.i.d. and sampled from $\mathcal{N}(0,1)$, and fixed values for $\delta$ and $\alpha$.\\
$\bullet$ {\emph{Case 2}}: The proposed G-RFPI algorithm where the learned sensing matrix $\bPhi$ and quantization thresholds vector $\bb$ are utilized, and the values for $\delta$ and $\alpha$ are fixed as in the previous case.\\
$\bullet$ {\emph{Case 3}}: The proposed one-bit LG-RFPI VAE method corresponsing to the iterations of the form \eqref{eq:31a}-\eqref{eq:31c}, with the learned $\bPhi$, $\bb$, $\{\delta_i\}_{i=0}^{L-1}$, and $\{\btau_i\}_{i=0}^{L-1}$.

Note that the focus of this part is on recovering the amplitude information of the underlying $K$-sparse signal by means of using arbitrary quantization thresholds. Although the RFPI method and the proposed L-RFPI VAE can only recover the normalized signal $\bx^d = \bx/\|\bx\|$, we further provide the performance of the L-RFPI method (that significantly outperforms the RFPI method) in recovering the amplitude information for comparison purposes. 
\begin{figure}
	\centering
	\includegraphics[width=0.3\textwidth]{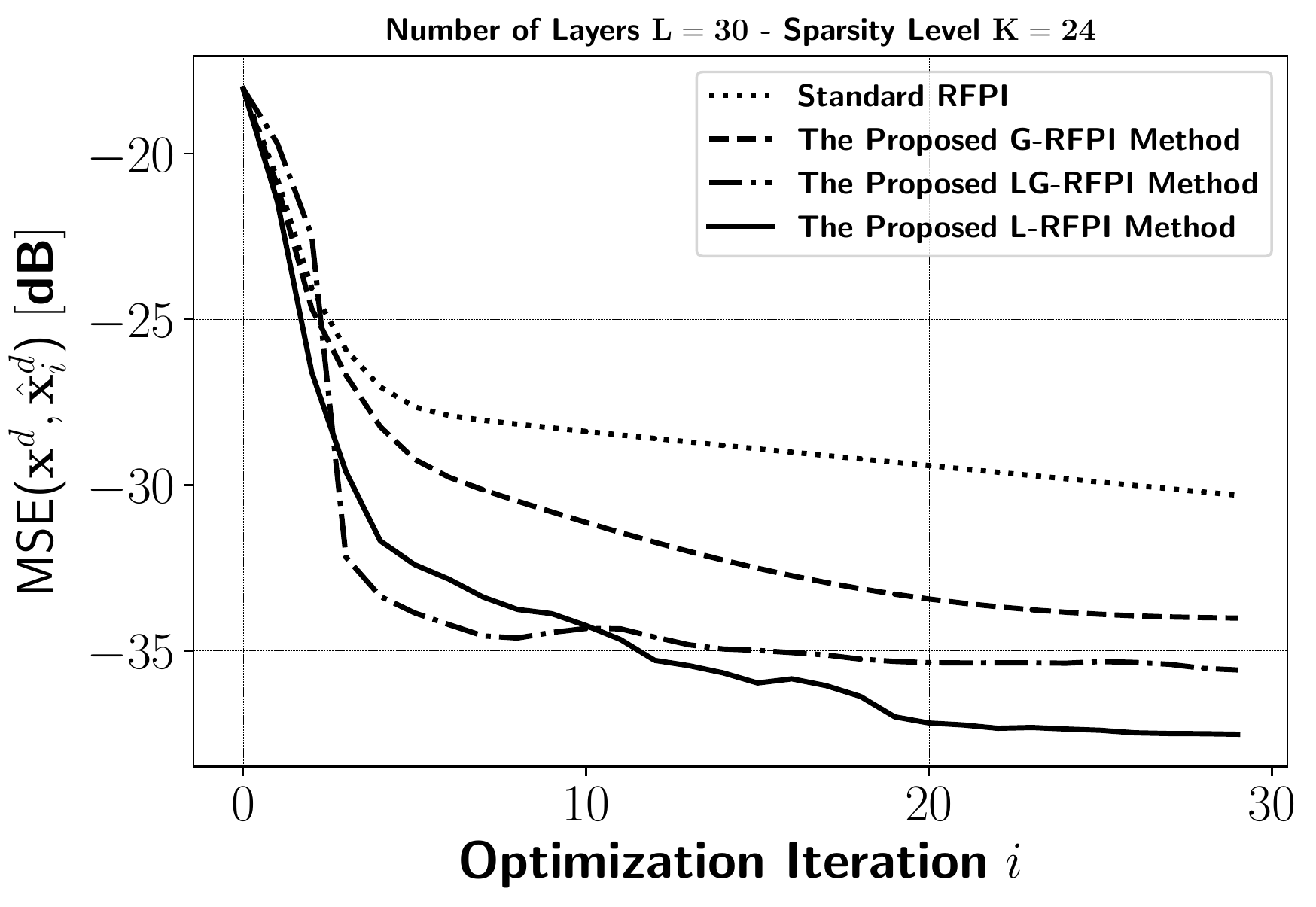}
	\caption{The performance of the proposed LG-RFPI VAE and the proposed G-RFPI method in recovering the normalized $K$-sparse signals for sparsity level $K=24$.}
	\label{fig:4}
\end{figure}
\begin{figure}
    \centering
    \subfigure[]{\includegraphics[width=0.24\textwidth]{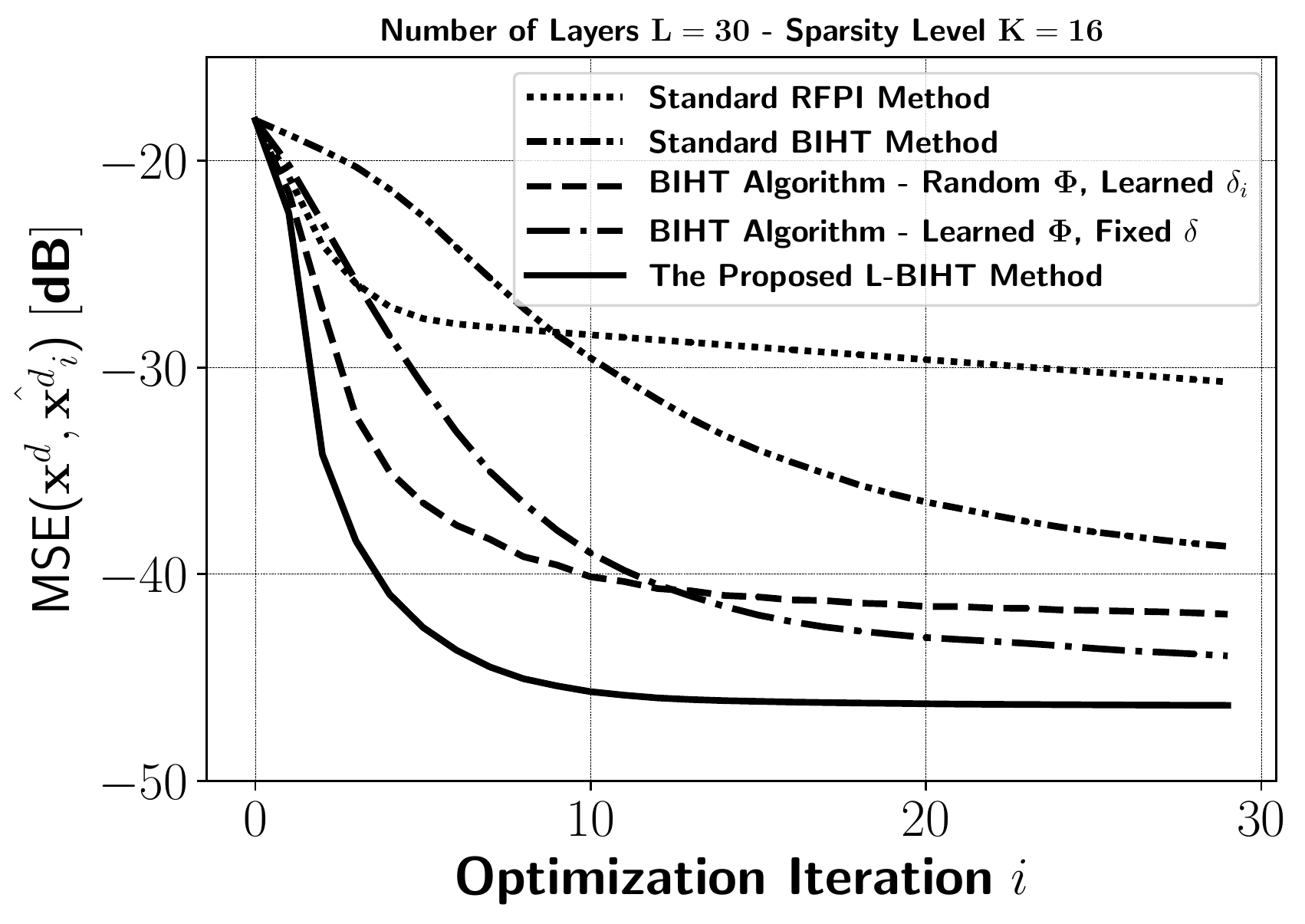}} 
    \subfigure[]{\includegraphics[width=0.24\textwidth]{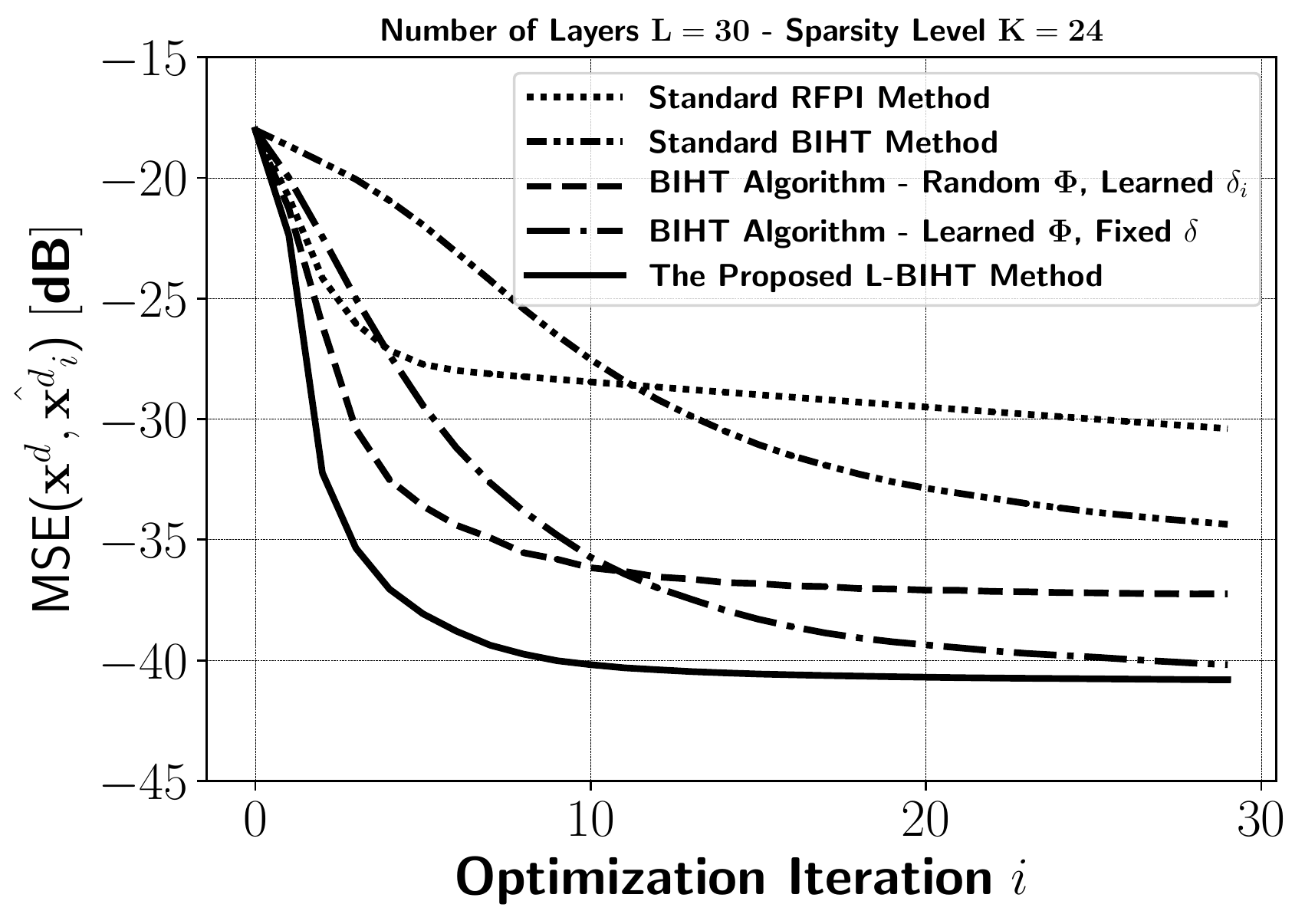}} 
    \caption{The performance of the proposed L-BIHT method compared to the base-line BIHT algorithm for sparsity levels: (a) $K=16$ and (b) $K=24$.}
    \label{fig:5}
\end{figure}
It can be observed from Fig. \ref{fig:3} that the proposed G-RFPI algorithm with randomly generated sensing matrix and quantization thresholds (Case 1) provides good accuracy in recovering the amplitude information of the true signal for sparsity levels $K\in\{8, 16,32,40\}$. This is in contrast to the RFPI algorithm and the corresponding L-RFPI VAE where the amplitude information is lost due to zero quantization thresholds. More precisely, the proposed G-RFPI algorithm outperforms the RFPI and the L-RFPI algorithm in terms of recovering the amplitude information of the signal. One can observe that even with a randomly generated quantization thresholds (i.e., without learning them), the proposed G-RFPI method achieve a significantly lower MSE in terms of recovering the amplitude information of the source signal as compared to the RFPI and the proposed L-RFPI method. Hence, the proposed G-RFPI method can be used as an stand-alone algorithm for one-bit compressive sensing settings with non-zero quantization thresholds, where both finding the direction of the source signal and the amplitude information is of great interest. Next, we explore the effect of learning the distribution-specific (data-driven) sensing matrix and the quantization thresholds (Case 2). It is evident from Fig. \ref{fig:3} that compared to the vanilla G-RFPI method, one can significantly achieve a lower MSE in terms of recovering the amplitude information by learning a proper sensing matrix and the quantization thresholds and utilizing them during the data-acquisition process. Finally, it can be seen from Fig. \ref{fig:3} that the proposed LG-RFPI VAE (Case 3) significantly outperforms its counterparts by achieving a much lower MSE very quickly. Moreover, the proposed LG-RFPI VAE shows strong generalization properties for unseen sparsity levels $K=\in\{8,40\}$ (see Fig. \ref{fig:3} (b) and (d)). The fact that such architectures show great performance in generalization is due to the model-driven nature of the proposed deep networks.

We conclude this part by comparing the performance of the proposed LG-RFPI, G-RFPI, and L-RFPI VAEs in recovering the normalized version of the signal $\bx$. Fig. \ref{fig:4} illustrates the MSE between the normalized source signal and the recovered signal versus number of iterations $i$, i.e. MSE($\bx^d$,$\hat{\bx}_{i}^d$), for a sparsity level of $K=24$. It can be observed from Fig. \ref{fig:3} that the proposed methods outperform the standard RFPI iterations and achieve a high accuracy in recovering $\bx^d$. Moreover, the proposed L-RFPI VAE shows a slightly better performance than that of the LG-RFPI method. This is presumably due to the fact that the L-RFPI iterations and the corresponding deep architecture are specifically designed and tuned for recovering the normalized source signal while the proposed G-RFPI and LG-RFPI algorithms are designed for recovering the amplitude information of the source signal. Nevertheless, the MSE difference between the LG-RFPI and L-RFPI methods in recovering $\bx^d$ is negligible, and hence, in a non-zero quantization thresholds setting, it is beneficial to use the proposed LG-RFPI VAE as it shows significant improvement in the performance of recovering the amplitude information while maintaining a high performance in recovering $\bx^\text{d}$ as well.

\noindent
\vspace{2pt}
\textbf{\underline{Performance of the proposed L-BIHT VAE}:}\\
In this part, we investigate the performance of the proposed L-BIHT VAE, and compare our results with the standard BIHT algorithm. Note that similar to the RFPI method and the proposed L-RFPI VAE, the BIHT algorithm considers $\bb=\mathbf{0}$ at the time of data acquisition. Hence, we investigate the performance of the proposed method in recovering the normalized source signal, i.e. $\bx^{\text{d}}$. In particular, we provide the simulation results for the following cases:\\
$\bullet$ {\emph{Case 1}}: The BIHT algorithm with a randomly generated sensing matrix whose elements are i.i.d. and sampled from $\mathcal{N}(0,1)$, and fixed value for $\delta$.\\
$\bullet$ {\emph{Case 2}}: The BIHT algorithm with a randomly generated $\bPhi$ (same as Case 1); however, learned gradient step-sizes $\delta_i$ are used at each iteration $i$.\\
$\bullet$ {\emph{Case 3}}: The BIHT algorithm where the learned $\bPhi$ is utilized and the value for the step-size $\delta$ is fixed as in Case 1.\\
$\bullet$ {\emph{Case 4}}: The proposed one-bit L-BIHT VAE method corresponding to the iterations of the form \eqref{eq:26a}-\eqref{eq:26b}, with the learned $\bPhi$ and $\{\delta_i\}_{i=0}^{L-1}$.
\begin{figure}
    \centering
    \subfigure[]{\includegraphics[width=0.24\textwidth]{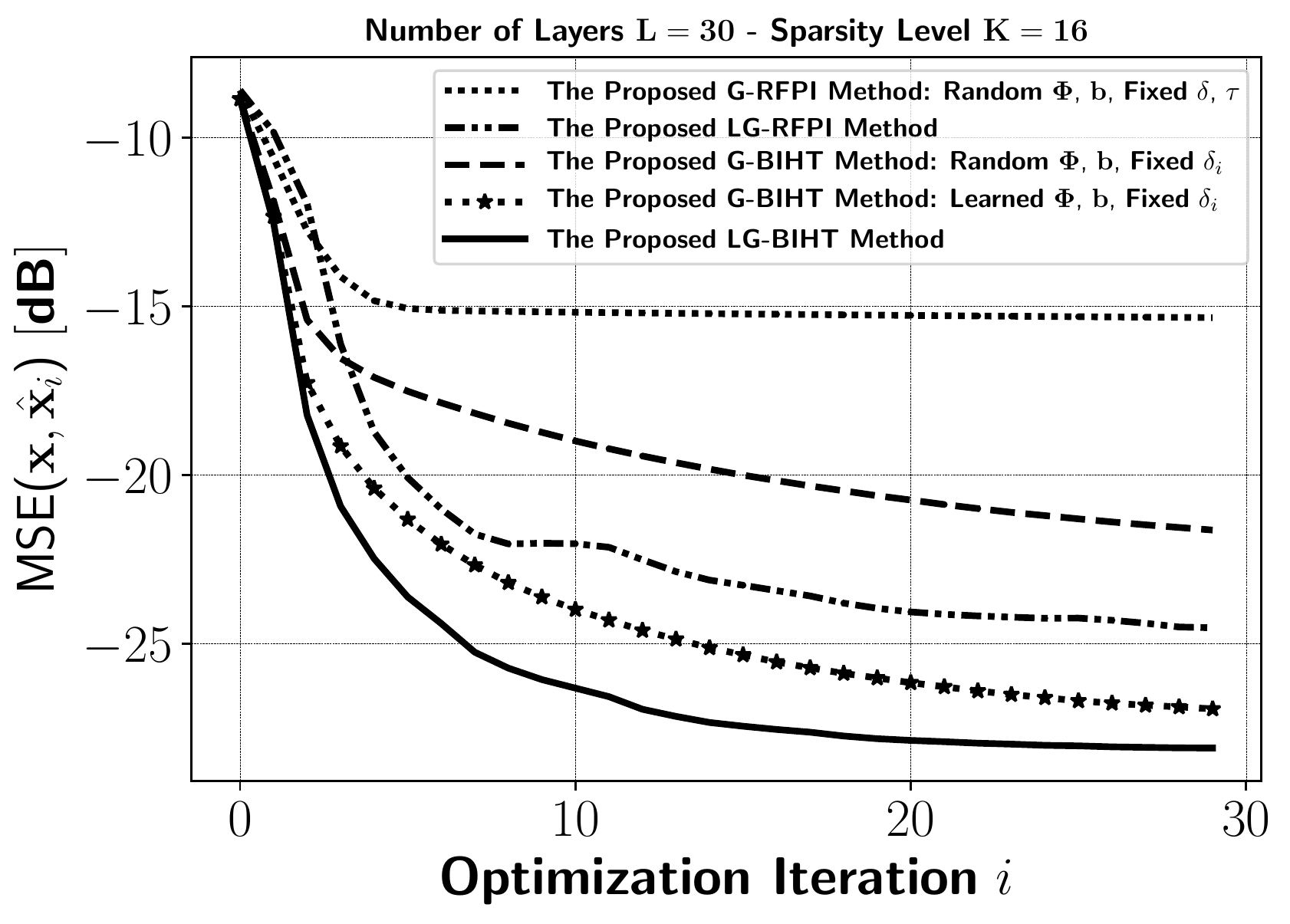}} 
    \subfigure[]{\includegraphics[width=0.24\textwidth]{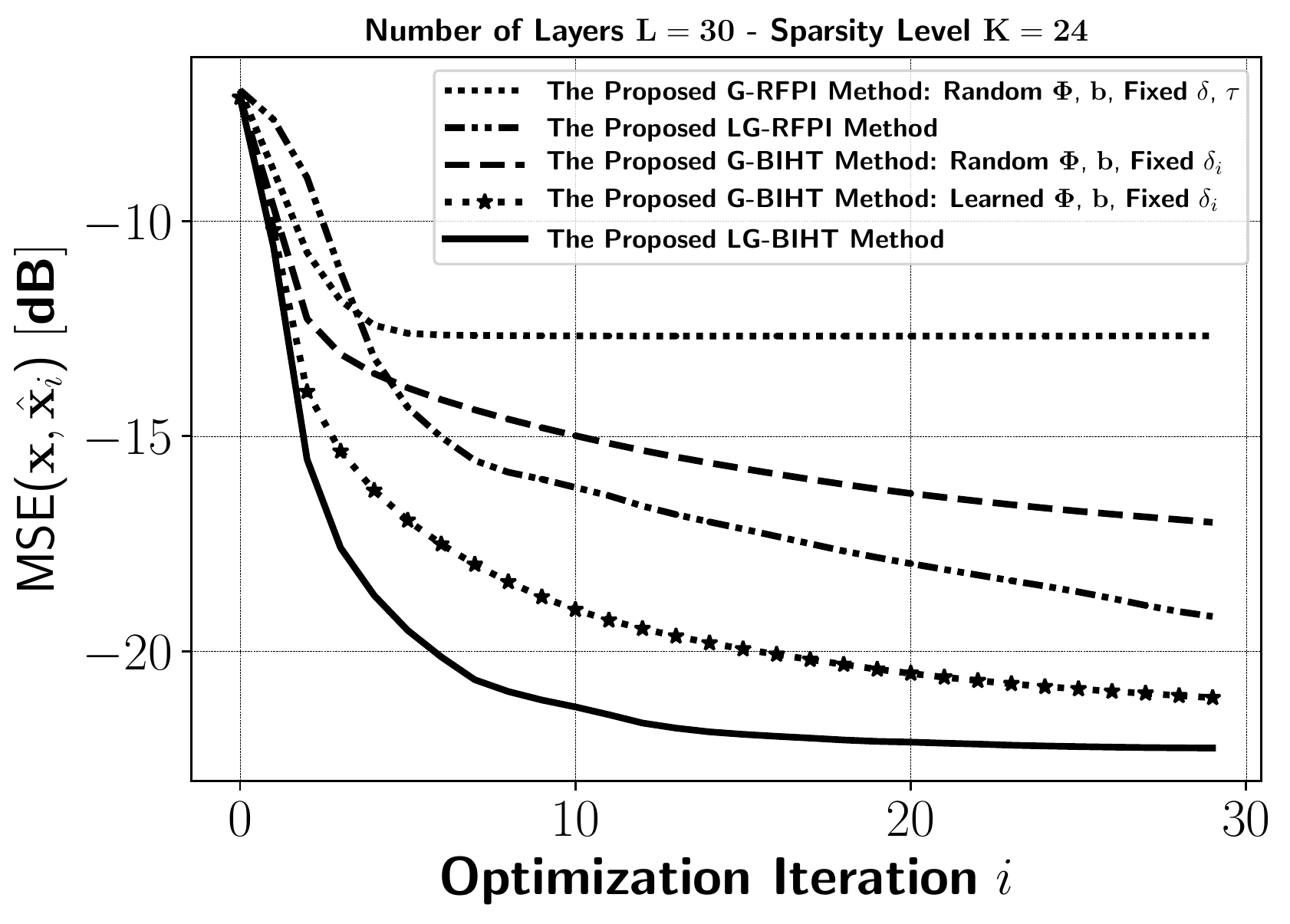}} 
    \caption{The performance of the proposed G-BIHT and the corresponding LG-BIHT VAE in recovering the amplitude information of the signal for sparsity levels: (a) $K=16$, and (b) $K=24$.}
    \label{fig:6}
\end{figure}
%

Fig. \ref{fig:5} demonstrates the MSE between normalized source signal $\bx^d$, and the recovered signal $\hat{\bx}^\text{d}_i$ versus the number of optimization iterations $i$, for signals with sparsity levels (a) $K=16$ and (b) $K=24$. Note that for learning the parameters of the proposed L-BIHT algorithm, we trained the corresponding deep architecture on the sparsity level $K=16$, and we check the generalization performance of the learned parameters for the case of $K=24$. It can be seen from Fig. \ref{fig:5} that in both cases of $K\in\{16,24\}$ the proposed L-BIHT algorithm demonstrates a significantly better performance than that of the standard BIHT algorithm (Case 1). Moreover, the effectiveness of the learned step-sizes $\{\delta_i\}_{i=0}^{L-1}$ (Case 2), and the learned sensing matrix $\bPhi$ (Case 3) compared to the base-line vanilla BIHT algorithm (Case 1) are evident. In particular, the learned step-sizes (Case 2) results in a fast descent while the learned $\bPhi$ (Case 3) leads to a lower MSE compared to Case 2. In addition, we provided the performance of the standard RFPI algorithm for comparison purposes. It can be seen from Fig. \ref{fig:5} that the BIHT algorithm with and without the learned parameters achieves a better accuracy in recovering the direction of the source signal compared to the RFPI method. Also, a comparison between Fig. \ref{fig:5} (a) and Fig. \ref{fig:2} (b) reveals the fact that the proposed L-BIHT VAE demonstrates a far better performance than that of the proposed L-RFPI VAE. This is due to the fact that the BIHT algorithm and the corresponding proposed L-BIHT VAE, exploits the knowledge of the sparsity level $K$ of the source signal (note the mapping function $\mathcal{H}_K$ used in \eqref{eq:26a} and \eqref{eqs:20}). One can further observe that even for the unseen case of $K=24$, the proposed method generalizes very well and maintains its accuracy. This is due the model-driven nature of the proposed L-BIHT VAE architecture. It is worth mentioning that it can be observed from Fig. \ref{fig:5} that the proposed L-BIHT method converges very fast (in 10 iterations), achieving a high accuracy---making it a great candidate for real-time applications. Of course, the trade-off between using the L-RFPI and L-BIHT is implicit in the knowledge of the sparsity level of the signal. For applications where $K$ is known beforehand, the proposed L-BIHT can be used in that it shows higher accuracy compared to the other methods. However, the L-RFPI methodology is more flexible as it does not require knowing the sparsity level of the signal a priori.\\
%
\begin{figure}
    \centering
    \subfigure[]{\includegraphics[width=0.24\textwidth]{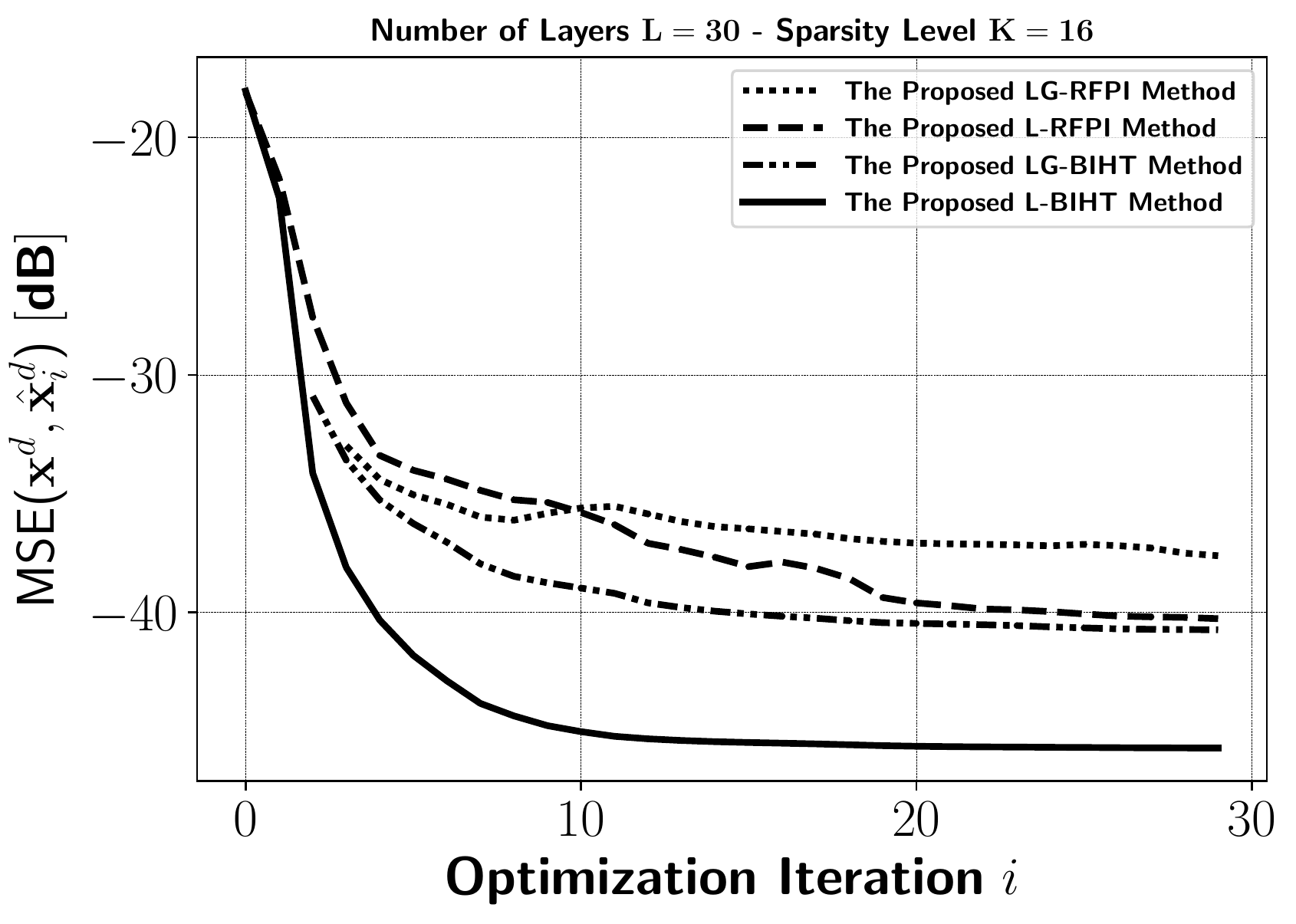}} 
    \subfigure[]{\includegraphics[width=0.24\textwidth]{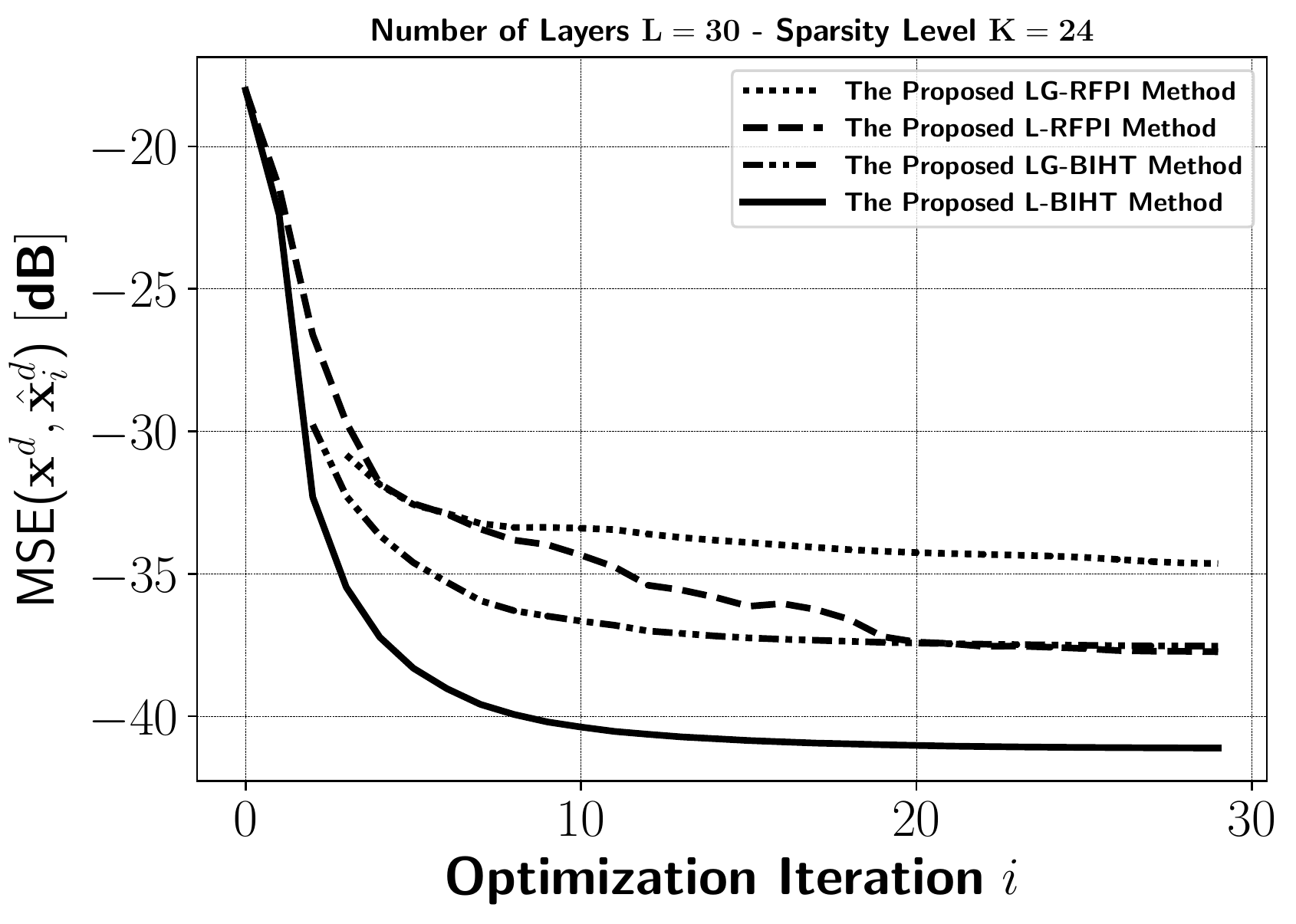}} 
    \caption{The performance of the proposed G-BIHT and the corresponding LG-BIHT VAE in recovering the normalized signal, i.e. $\bx^\text{d}$, for sparsity levels: (a) $K=16$, and (b) $K=24$.}
    \label{fig:7}
\end{figure}
\noindent
\vspace{2pt}
\textbf{\underline{Performance of the proposed LG-BIHT VAE}:}\\
Finally, we investigate the performance of the proposed G-BIHT method (see Eqs. \eqref{eqq:21a}-\eqref{eqq:21b}) and the corresponding one-bit compressive LG-BIHT VAE (see Eqs. \eqref{eq:25a}-\eqref{eq:25b}) that are specifically designed to handle non-zero quantization thresholds $\bb$. In particular, we are interested in evaluating the performance of the proposed methods in recovering the amplitude information of the source $K$-sparse signal. Hence, for this part, we check the MSE between the true signal $\bx$, and the recovered signal $\hat{\bx}_i$ from the G-BIHT and LG-BIHT methods for each iteration $i$. In addition, we provide the results for recovering the direction of the source signal $\bx^\text{d}$ as well. Specifically, we provide the simulation results for the following cases:\\
$\bullet$ {\emph{Case 1}}: The proposed G-BIHT algorithm with a randomly generated sensing matrix and quantization thresholds vector where the elements of both are i.i.d. and sampled from $\mathcal{N}(0,1)$, and fixed value for $\{\delta_i\}_{i=0}^{L-1}$.\\
$\bullet$ {\emph{Case 2}}: The proposed G-BIHT algorithm, where the learned sensing matrix $\bPhi$ and quantization thresholds $\bb$ are utilized and the values for $\{\delta_i\}_{i=0}^{L-1}$ are fixed as in the previous case.\\
$\bullet$ {\emph{Case 3}}: The proposed one-bit LG-BIHT VAE method corresponding to the iterations of the form \eqref{eq:25a}-\eqref{eq:25b}, with learned $\bPhi$, $\bb$ and $\{\delta_i\}_{i=0}^{L-1}$.

Fig. \ref{fig:6} illustrates the MSE between the true signal $\bx$ and the recovered signal $\hat{\bx}_i$ versus optimization iteration $i$ for sparsity levels (a) $K=16$ and (b) $K=24$. We further provide the numerical results for the proposed LG-RFPI VAE and the proposed G-RFPI iterations for comparison. It can be seen from Fig. \ref{fig:6} that the proposed G-BIHT algorithm with randomly generated latent-variables (Case 1) significantly outperforms its G-RFPI counterpart, and achieves a high accuracy very quickly. On the other hand, the proposed LG-RFPI still achieves a lower MSE compared to the vanilla G-RFPI method. In addition, a comparison between the performance of the proposed G-BIHT algorithm with learned $\bPhi$ and $\bb$ (Case 2) and the proposed LG-RFPI VAE and vanilla G-BIHT (Case 1) reveals the effectiveness of the learned parameters and the power of the proposed G-BIHT algorithm. Namely, by utilizing only the learned $\bPhi$ and $\bb$ and by using a fixed step size for the G-BIHT algorithm, one can achieve a superior performance than that of the LG-RFPI (where all of the learned variables are in use) and the vanilla G-BIHT method. Finally, it can be observed from \ref{fig:5}(a)-(b) that the proposed LG-BIHT algorithm (Case 3) significantly outperforms the other methods as it achieves a much lower MSE very quickly, specifically, compared to the proposed LG-RFPI VAE. The superior performance of the LG-BIHT algorithm and the corresponding LG-BIHT VAE is due the fact that we are exploiting the knowledge of the sparsity level $K$ present in the signal. As discussed before, if the sparsity level is known a priori, it is beneficial to use either the G-BIHT algorithm (when one do not wish to perform any learning) or the proposed LG-BIHT methodology. It is worth mentioning that similar to the previously investigated methods, the proposed LG-BIHT generalizes very well for $K=24$ (see Fig. \ref{fig:6}(b)) even though the sparsity level was not revealed to the network during the training phase.

Fig. \ref{fig:7} demonstrate the MSE between the direction of the source signal, i.e. $\bx^{\text{d}}$, and the recovered direction $\hat{\bx}^{\text{d}}$ versus optimization iteration $i$, for sparsity levels of (a) $K=16$ and (b) $K=24$. It can be seen from Fig. \ref{fig:7} that the proposed LG-BIHT method outperforms both the LG-RFPI method, and furthermore, it achieves a similar MSE to that of the proposed L-RFPI method. However, the convergence of LG-BIHT is much faster than that of the L-RFPI method. Furthermore, the proposed L-BIHT algorithm still achieves a superior performance than that of the other methods both in terms of convergence speed and accuracy. This is presumably due to the fact that the L-BIHT method is specifically designed and learned to have a high accuracy in finding normalized true signal $\bx^{\text{d}}$. 
\section{Conclusion}
In this paper, we considered the problem of one-bit compressive sensing and proposed a novel hybrid \emph{model-driven} and \emph{data-driven} variational autoencoding scheme that allows us to \emph{jointly} learn the parameters of the measurment module (i.e., the sensing matrix and the quantization thresholds) and the latent-variables of the decoder (estimator) function, based on the underlying distribution of the data. In broad terms, we proposed a novel methodology that combines the traditional compressive sensing techniques with model-based deep learning---resulting in interpretable deep architectures for the problem of one-bit compressive sensing. In addition, the proposed method can handle the recovery of the amplitude information of the signal using the learned and optimized quantization thresholds. Our simulation results demonstrated that the proposed hybrid methodology is superior to the state-of-the-art methods for the problem of one-bit CS in terms of both computional efficiency and accuracy.
\bibliographystyle{IEEEbib}
\balance

\bibliography{refs}


\end{document}